\input harvmac
%\input pdftexconfig
%\input graphicx
%\input epsf
%\def\Title#1#2{\rightline{#1}\ifx\answ\bigans\nopagenumbers\pageno0\vskip1in
%\else\pageno1\vskip.8in\fi \centerline{\titlefont #2}\vskip .5in}
%%
%\font\titlerm=cmRI0 scaled\magstep3 \font\titlerms=cmr7 scaled\magstep3
%\font\titlermss=cmr5 scaled\magstep3 \font\titlei=cmmi10 scaled\magstep3
%\font\titleis=cmmi7 scaled\magstep3 \font\titleiss=cmmi5 scaled\magstep3
%\font\titlesy=cmsy10 scaled\magstep3 \font\titlesys=cmsy7 scaled\magstep3
%\font\titlesyss=cmsy5 scaled\magstep3 \font\titleit=cmti10 scaled\magstep3
%%%%%%%%%%%%%%%%%%
%
% Figure macros, SBG 5/93
%
%\ifx\epsfbox\UnDeFiNeD\message{(NO epsf.tex, FIGURES WILL BE IGNORED)}
%\def\figin#1{\vskip2in}% blank space instead
%\else\message{(FIGURES WILL BE INCLUDED)}\def\figin#1{#1}%\epsfverbosetrue
%\fi
%\def\Fig#1{Fig.~\the\figno\xdef#1{Fig.~\the\figno}\global\advance\figno
% by1}
%%
%%  ifig   usage:
%%
%%         \ifig\figlabel{caption}{figfile}{vsize}
%%
%% where vsize is the desired vertical size of the figure in truein
%%
%\def\ifig#1#2#3#4{
%\goodbreak\midinsert
%\figin{\centerline{\epsfysize=#4truein\epsfbox{#3}}}
%\narrower\narrower\noindent{\footnotefont
%{\bf #1:}  #2\par}
%\endinsert
%}
%\def\ifigx#1#2#3#4{
%\goodbreak\midinsert
%\figin{\centerline{\epsfxsize=#4truein\epsfbox{#3}}}
%\narrower\narrower\noindent{\footnotefont
%{\bf #1:}  #2\par}
%\endinsert
%}
%%%%%%%%%%%%%%%%%%
%
% Figure macros, SBG 3/03
%
\ifx\includegraphics\UnDeFiNeD\message{(NO graphicx.tex, FIGURES WILL BE IGNORED)}
\def\figin#1{\vskip2in}% blank space instead
\else\message{(FIGURES WILL BE INCLUDED)}\def\figin#1{#1}
\fi
\def\Fig#1{Fig.~\the\figno\xdef#1{Fig.~\the\figno}\global\advance\figno
 by1}
%
%  Ifig   usage:
%
%         \Ifig{\Fig\figlabel}{caption}{figfile}{hsize}
%
% where vsize is the desired vertical size of the figure in truein
%

%
%defs
%
\font\ticp=cmcsc10
\def\undertext#1{$\underline{\smash{\hbox{#1}}}$}

\def\hf{{1\over 2}}
\def\calo{{\cal O}}
\def\pip { [ \! ]   }
\def\calv{{\cal V}}
\def\cohat{{\hat \calo}}
\def\cotilde{{\tilde \calo}}
\def\calh{{\cal H}}
\def\cald{{\cal D}}
\def\calV{{\cal V}}
\def\mpl{{M_{p}}}
\def\rtg{{\sqrt{-g}}}

\def\subsubsec#1{\noindent{\undertext { #1}}}
\def\mthsu{\mathsurround=0pt  }
\def\leftrightarrowfill{$\mthsu \mathord\leftarrow\mkern-6mu\cleaders
  \hbox{$\mkern-2mu \mathord- \mkern-2mu$}\hfill
  \mkern-6mu\mathord\rightarrow$}
 \def\overleftrightarrow#1{\vbox{\ialign{##\crcr\leftrightarrowfill\crcr\noalign{\kern-1pt\nointerlineskip}$\hfil\displaystyle{#1}\hfil$\crcr}}}
\overfullrule=0pt
%
%refs
%

%

\lref\Ambjorn{
  J.~Ambjorn and K.~N.~Anagnostopoulos,
  ``Quantum geometry of 2D gravity coupled to unitary matter,''
  Nucl.\ Phys.\ B {\bf 497}, 445 (1997)
  [arXiv:hep-lat/9701006].
  %%CITATION = HEP-LAT 9701006;%%
}
\lref\SumOver{M.~P.~Reisenberger and C.~Rovelli,
  ``Sum over surfaces form of loop quantum gravity,''
  Phys.\ Rev.\ D {\bf 56}, 3490 (1997)
  [arXiv:gr-qc/9612035].
  %%CITATION = GR-QC 9612035;%%
}

\lref\Projector{  C.~Rovelli,
  ``The projector on physical states in loop quantum gravity,''
  Phys.\ Rev.\ D {\bf 59}, 104015 (1999)
  [arXiv:gr-qc/9806121].
  %%CITATION = GR-QC 9806121;%%
}

\lref\Partial{  C.~Rovelli,
  ``Partial observables,''
  Phys.\ Rev.\ D {\bf 65}, 124013 (2002)
  [arXiv:gr-qc/0110035].
  %%CITATION = GR-QC 0110035;%%
}

\lref\StiffStars{ T.~Banks, W.~Fischler, A.~Kashani-Poor, R.~McNees and S.~Paban,
  ``Entropy of the stiffest stars,''
  Class.\ Quant.\ Grav.\  {\bf 19}, 4717 (2002)
  [arXiv:hep-th/0206096].
  %%CITATION = HEP-TH 0206096;%%
}

\lref\Smolin{L.~Smolin,
  ``Time, measurement and information loss in quantum cosmology,''
  arXiv:gr-qc/9301016.
  %%CITATION = GR-QC 9301016;%%
}

\lref\TW{  N.~C.~Tsamis and R.~P.~Woodard,
  ``Physical Green's Functions In Quantum Gravity,''
  Annals Phys.\  {\bf 215}, 96 (1992).
  %%CITATION = APNYA,215,96;%%
}

\lref\Hepp{ K. Hepp, ``Quantum theory of measurement and macroscopic observables," Helv. Phys. Acta. {\bf 45} 237 (1972).}

\lref\Bell{J. S. Bell, ``On wave packet reduction in the Coleman-Hepp model," Helv. Phys. Acta. {\bf 48} 93 (1975).}

\lref\Dirac{
P. A. M. Dirac,  {\it Lectures on Quantum Mechanics}, (Belfor Graduate School of Science, Yeshiva University, New York, 1964). }

\lref\GiddingsWV{
  S.~B.~Giddings and A.~Strominger,
  ``Baby Universes, Third Quantization And The Cosmological Constant,''
  Nucl.\ Phys.\ B {\bf 321}, 481 (1989).
  %%CITATION = NUPHA,B321,481;%%
}

\lref\BanksJE{
  T.~Banks,
  ``Prolegomena To A Theory Of Bifurcating Universes: A Nonlocal Solution To
  The Cosmological Constant Problem Or Little Lambda Goes Back To The Future,''
  Nucl.\ Phys.\ B {\bf 309}, 493 (1988).
  %%CITATION = NUPHA,B309,493;%%
}

\lref\BryceI{
B. DeWitt, ``The Quantization of Geometry'', in
{\it Gravitation: An Introduction to Current Research},  ed. Witten L (New
York, Wiley, 1962).}

\lref\BryceII{
  B.~S.~DeWitt,
  ``Quantum Theory Of Gravity. 1. The Canonical Theory,''
  Phys.\ Rev.\  {\bf 160}, 1113 (1967).
  %%CITATION = PHRVA,160,1113;%%
}

\lref\RI{C. Rovelli, in Conceptual Problems of Quantum Gravity ed. by
Ashtekar A and Stachel J (Boston: Birkh\"auser, 1991) 141.}

\lref\RII{
  C.~Rovelli,
  ``Quantum Mechanics Without Time: A Model,''
  Phys.\ Rev.\ D {\bf 42}, 2638 (1990).}
\lref\CKN{
  A.~G.~Cohen, D.~B.~Kaplan and A.~E.~Nelson,
  ``Effective field theory, black holes, and the cosmological constant,''
  Phys.\ Rev.\ Lett.\  {\bf 82}, 4971 (1999)
  [arXiv:hep-th/9803132].
  %%CITATION = HEP-TH 9803132;%%
}
\lref\Hawkunc{
  S.~W.~Hawking,
  ``Breakdown Of Predictability In Gravitational Collapse,''
  Phys.\ Rev.\ D {\bf 14}, 2460 (1976).
  %%CITATION = PHRVA,D14,2460;%%
}

\lref\RIII{C. Rovelli, ``Time in quantum gravity: An hypothesis,'' Phys. Rev. D {\bf 43} 442 (1991).}
\lref\GiLione{
  S.~B.~Giddings and M.~Lippert,
  ``Precursors, black holes, and a locality bound,''
  Phys.\ Rev.\ D {\bf 65}, 024006 (2001)
  [arXiv:hep-th/0103231].
  %%CITATION = HEP-TH 0103231;%%
}
\lref\GiLitwo{
  S.~B.~Giddings and M.~Lippert,
  ``The information paradox and the locality bound,''
  Phys.\ Rev.\ D {\bf 69}, 124019 (2004)
  [arXiv:hep-th/0402073].
  %%CITATION = HEP-TH 0402073;%%
}

\lref\SBGBSM{
  S.~B.~Giddings,
  ``The boundary S-matrix and the AdS to CFT dictionary,''
  Phys.\ Rev.\ Lett.\  {\bf 83}, 2707 (1999)
  [arXiv:hep-th/9903048].
  %%CITATION = HEP-TH 9903048;%%
}

\lref\HarMar { J.~B.~Hartle and D.~Marolf,
  ``Comparing formulations of generalized quantum mechanics for
  reparametrization-invariant systems,''
  Phys.\ Rev.\ D {\bf 56}, 6247 (1997)
  [arXiv:gr-qc/9703021].
  %%CITATION = GR-QC 9703021;%%
}

\lref\RIV{C. Rovelli, ``What is observable in classical and quantum gravity?'' Class. Quantum Grav. {\bf 8} 297  (1991).}

\lref\QORD{
  D.~Marolf,
  ``Quantum observables and recollapsing dynamics,''
  Class.\ Quant.\ Grav.\  {\bf 12}, 1199 (1995)
  [arXiv:gr-qc/9404053].
  %%CITATION = GR-QC 9404053;%%
}

\lref\ToF{
  N.~Grot, C.~Rovelli and R.~S.~Tate,
 ``Time-of-arrival in quantum mechanics,''
  Phys.\ Rev.\ A {\bf 54}, 4676 (1996)
  [arXiv:quant-ph/9603021].
  %%CITATION = QUANT-PH 9603021;%%
}

\lref\BIX{
  D.~Marolf,
  ``Observables and a Hilbert space for Bianchi IX,''
  Class.\ Quant.\ Grav.\  {\bf 12}, 1441 (1995)
  [arXiv:gr-qc/9409049].
  %%CITATION = GR-QC 9409049;%%
} \lref\EinsDov{A. Einstein, {\sl Relativity: the special and
general theory} (Dover, New York, 1961).}

\lref\MarT{
  D.~Marolf,
  ``Almost ideal clocks in quantum cosmology: A Brief derivation of time,''
  Class.\ Quant.\ Grav.\  {\bf 12}, 2469 (1995)
  [arXiv:gr-qc/9412016].
  %%CITATION = GR-QC 9412016;%%
}

\lref\Higuchi{
  A.~Higuchi,
  ``Quantum linearization instabilities of de Sitter space-time. 2,''
  Class.\ Quant.\ Grav.\  {\bf 8}, 1983 (1991).
  %%CITATION = CQGRD,8,1983;%%
}

\lref\Landsman{
  N.~P.~Landsman,
  ``Rieffel induction as generalized quantum Marsden-Weinstein reduction,''
  arXiv:hep-th/9305088.
  %%CITATION = HEP-TH 9305088;%%
}

\lref\HaQMC{ J.~B.~Hartle,  ``The Quantum Mechanics Of Cosmology,'' in {\it Quantum Cosmology and Baby Universes: Proceedings of the 1989 Jerusalem Winter School for Theoretical Physics,} eds. S. Coleman, J.B. Hartle, T. Piran, and S. Weinberg, World Scientific, Singapore (1991) pp. 65-157. }

\lref\PageUC{
  D.~N.~Page and W.~K.~Wootters,
  ``Evolution Without Evolution: Dynamics Described By Stationary
  Observables,''
  Phys.\ Rev.\ D {\bf 27}, 2885 (1983).
  %%CITATION = PHRVA,D27,2885;%%
}

\lref\ALMMT{
  A.~Ashtekar, J.~Lewandowski, D.~Marolf, J.~Mourao and T.~Thiemann,
  ``Quantization of diffeomorphism invariant theories of connections with local
  degrees of freedom,''
  J.\ Math.\ Phys.\  {\bf 36}, 6456 (1995)
  [arXiv:gr-qc/9504018].
  %%CITATION = GR-QC 9504018;%%
}

\lref\MarOne{
  D.~Marolf,
  ``Refined algebraic quantization: Systems with a single constraint,''
  arXiv:gr-qc/9508015.
  %%CITATION = GR-QC 9508015;%%
}

\lref\WhereAreWe{
  D.~Marolf,
  ``Group averaging and refined algebraic quantization: Where are we now?,''
  arXiv:gr-qc/0011112.
  %%CITATION = GR-QC 0011112;%%
}

\lref\OS{
  O.~Y.~Shvedov,
  ``Refined algebraic quantization of constrained systems with structure
  functions,''
  arXiv:hep-th/0107064.
  %%CITATION = HEP-TH 0107064;%%
}

\lref\OSII{
  O.~Y.~Shvedov,
  ``On correspondence of BRST-BFV, Dirac and refined algebraic  quantizations
  of constrained systems,''
  Annals Phys.\  {\bf 302}, 2 (2002)
  [arXiv:hep-th/0111270].
  %%CITATION = HEP-TH 0111270;%%
}

\lref\BanksCG{
  T.~Banks,
  ``Some Thoughts on the Quantum Theory of de Sitter Space,''
  arXiv:astro-ph/0305037.
  %%CITATION = ASTRO-PH 0305037;%%
}

\lref\BanksWR{
  T.~Banks, W.~Fischler and S.~Paban,
  ``Recurrent nightmares?: Measurement theory in de Sitter space,''
  JHEP {\bf 0212}, 062 (2002)
  [arXiv:hep-th/0210160].
  %%CITATION = HEP-TH 0210160;%%
}

\lref\AstrodS{
  A.~Strominger,
  ``The dS/CFT correspondence,''
  JHEP {\bf 0110}, 034 (2001)
  [arXiv:hep-th/0106113].
  %%CITATION = HEP-TH 0106113;%%
}

\lref\Page{
  D.~N.~Page and W.~K.~Wootters,
  ``Evolution Without Evolution: Dynamics Described By Stationary
  Observables,''
  Phys.\ Rev.\ D {\bf 27}, 2885 (1983).
  %%CITATION = PHRVA,D27,2885;%%
}

\lref\WittdS{
  E.~Witten,
  ``Quantum gravity in de Sitter space,''
  arXiv:hep-th/0106109.
  %%CITATION = HEP-TH 0106109;%%
}

\lref\Kuchar{
K. Kucha\v{r}, ``Time and Interpretations of Quantum Gravity,'' in {\it Proceedings of the 4th Canadian Conference on General Relativity and Relativistic Astrophysics} ed. Kunstatter G et. al. (New Jersey: World Scientific, 1992).}

\lref\ATU{
  A.~Ashtekar, R.~Tate and C.~Uggla,
  ``Minisuperspaces: Observables and quantization,''
  Int.\ J.\ Mod.\ Phys.\ D {\bf 2}, 15 (1993)
  [arXiv:gr-qc/9302027].
  %%CITATION = GR-QC 9302027;%%
}

\lref\HartleCos{
  J.~B.~Hartle,
  ``Space-time quantum mechanics and the quantum mechanics of space-time,'' in
  {\it Gravitation and Quantizations: Proceedings of the 1992 Les Houches Summer School,} ed. by B. Julia and J. Zinn-Justin, Les Houches Summer School Proceedings v LVII, North Holland, Amsterdam (1995); gr-qc/9304006.  %%CITATION = GR-QC 9304006;%%
}

\lref\Dolby{
  C.~E.~Dolby,
  ``The conditional probability interpretation of the Hamiltonian
  constraint,''
  arXiv:gr-qc/0406034.
  %%CITATION = GR-QC 0406034;%%
}

\lref\Omnes{ R. Omnes,  {\it Understanding Quantum Mechanics}
(Princeton University Press, Princeton, 1999). }

\lref\Joos{E. Joos, E.,  {\it Decoherence and the Appearance of a Classical World in Quantum Theory}, 2nd edition (Springer, Berlin, 2003).}

\lref\ZI{  W. H. Zurek,  ``Decoherence and the transition from quantum to classical Ñ REVISITED", arXiv:quant-ph/0306072.}

\lref\ZII{ W. H. Zurek,  ``Decoherence, einselection, and the quantum origins of the classical", Rev.  Mod. Phys, {\bf 75} 715 (2003) [arXiv:quant-ph/0105127.]}

\lref\Schloss{M. Schlosshauer,  ``Decoherence, the Measurement
Problem, and Interpretations of Quantum Mechanics", Rev. of Mod.
Phys, {\bf 76} (2005) 1267 [arXiv:quant-ph/0312059]. }

\lref\tHooftGX{
  G.~'t Hooft,
``Dimensional reduction in quantum gravity,''
  arXiv:gr-qc/9310026.
  %%CITATION = GR-QC 9310026;%%
}
\lref\Bankslittle{
T.~Banks,
``Cosmological breaking of supersymmetry or little Lambda goes back to  the future. II,''
arXiv:hep-th/0007146.
%%CITATION = HEP-TH 0007146;%%
}
\lref\Fischler{W. Fischler, unpublished (2000);
W. Fischler, ``Taking de Sitter seriously," Talk given at
Role of Scaling Laws in Physics and Biology (Celebrating
the 60th Birthday of Geoffrey West), Santa Fe, Dec. 2000.}

\lref\GMII{
  D.~Giulini and D.~Marolf,
  ``A uniqueness theorem for constraint quantization,''
  Class.\ Quant.\ Grav.\  {\bf 16}, 2489 (1999)
  [arXiv:gr-qc/9902045].
  %%CITATION = GR-QC 9902045;%%
}

\lref\AD{
  A.~Gomberoff and D.~Marolf,
  ``On group averaging for SO(n,1),''
  Int.\ J.\ Mod.\ Phys.\ D {\bf 8}, 519 (1999)
  [arXiv:gr-qc/9902069].
  %%CITATION = GR-QC 9902069;%%
}

\lref\JLI{
  J.~Louko and C.~Rovelli,
  ``Refined algebraic quantization in the oscillator representation of
  SL(2,R),''
  J.\ Math.\ Phys.\  {\bf 41}, 132 (2000)
  [arXiv:gr-qc/9907004].
  %%CITATION = GR-QC 9907004;%%
}

\lref\JLII{
  J.~Louko and A.~Molgado,
  ``Superselection sectors in the Ashtekar-Horowitz-Boulware model,''
  Class.\ Quant.\ Grav.\  {\bf 22}, 4007 (2005)
  [arXiv:gr-qc/0505097].
  %%CITATION = GR-QC 0505097;%%
}

\lref\MPI{
  D.~Marolf,
  ``Path Integrals and Instantons in Quantum Gravity,''
  Phys.\ Rev.\ D {\bf 53}, 6979 (1996)
  [arXiv:gr-qc/9602019].
  %%CITATION = GR-QC 9602019;%%
}

\lref\HarHal{
  J.~J.~Halliwell and J.~B.~Hartle,
  ``Wave Functions Constructed From An Invariant Sum Over Histories Satisfy
  Constraints,''
  Phys.\ Rev.\ D {\bf 43}, 1170 (1991).
  %%CITATION = PHRVA,D43,1170;%%
}

\lref\JLIII{
  J.~Louko and A.~Molgado,
  ``Group averaging in the (p,q) oscillator representation of SL(2,R),''
  J.\ Math.\ Phys.\  {\bf 45}, 1919 (2004)
  [arXiv:gr-qc/0312014].
  %%CITATION = GR-QC 0312014;%%
}

\lref\vonN {J. von Neumann,
{\it Mathematical Foundations of Quantum Mechanics} (Princeton University Press: Princeton 1955).}

\lref\IKI { C.~J.~Isham and K.~V.~Kuchar,
  ``Representations Of Space-Time Diffeomorphisms. 1. Canonical Parametrized
  Field Theories,''
  Annals Phys.\  {\bf 164}, 288 (1985).}

\lref\IKII{
  C.~J.~Isham and K.~V.~Kuchar,
  ``Representations Of Space-Time Diffeomorphisms. 2. Canonical
  Geometrodynamics,''
  Annals Phys.\  {\bf 164}, 316 (1985).
  %%CITATION = APNYA,164,316;%%
}

\lref\Rubakov {  V.~G.~Lapchinsky and V.~A.~Rubakov,
  ``Canonical Quantization Of Gravity And Quantum Field Theory In Curved
  Space-Time,''
  Acta Phys.\ Polon.\ B {\bf 10}, 1041 (1979).
  %%CITATION = APPOA,B10,1041;%%
}

\lref\BanksTCP {  T.~Banks,
  ``T C P, Quantum Gravity, The Cosmological Constant And All That..,''
  Nucl.\ Phys.\ B {\bf 249}, 332 (1985).
  %%CITATION = NUPHA,B249,332;%%
}

\lref\Kiefer {
  C.~Kiefer and T.~P.~Singh,
  ``Quantum Gravitational Corrections To The Functional Schrodinger Equation,''
  Phys.\ Rev.\ D {\bf 44}, 1067 (1991).
  %%CITATION = PHRVA,D44,1067;%%
}
\lref\HaIt{
  A.~Hashimoto and N.~Itzhaki,
  ``Observables of string field theory,''
  JHEP {\bf 0201}, 028 (2002)
  [arXiv:hep-th/0111092].
  %%CITATION = HEP-TH 0111092;%%
}
\lref\GHI{
  D.~J.~Gross, A.~Hashimoto and N.~Itzhaki,
  ``Observables of non-commutative gauge theories,''
  Adv.\ Theor.\ Math.\ Phys.\  {\bf 4}, 893 (2000)
  [arXiv:hep-th/0008075].
  %%CITATION = HEP-TH 0008075;%%
}

\lref\Pons{J.M.Pons and D.C.Salisbury,
``The issue of time in generally covariant theories and the
Komar-Bergmann approach to observables in general relativity,''
  Phys.\ Rev.\ D {\bf 71}, 124012 (2005)
  [arXiv:gr-qc/0503013].}

\lref\Schwing{J. S. Schwinger, ``On The GreenÕs Functions Of Quantized Fields. 1," Proc. Nat. Acad. Sci. {\bf 37}, 452 (1951).}

\lref\BryceBook{ B. S. DeWitt, {\it Dynamical Theory of Groups and
Fields}, (Gordon and Breach, New York, 1965); B. S. DeWitt, in
{\it Relativity, Groups, and Topology II: Les Houches 1983,
Session XL}, (North-Holland, Amsterdam, 1984) edited by B. S.
DeWitt and R. Stora; B. S. DeWitt, The Global Approach to Quantum
Field Theory, (Oxford University Press, 2003). }

\lref\Hamber{H. Hamber,  ``Simplicial Quantum Gravity,'' in {\it Gauge Theories, Critical Phenomena, and Random Systems}, Proceedings of the 1984 Les Houches Summer School, ed. by K. Osterwalder and R. Stora (North Holland, 1986) and
`` Invariant Correlations in Simplicial  Gravity,'' {\it Phys. Rev D},
{\bf 50}, 3932 (1994). }
\lref\GambiniYJ{
  R.~Gambini, R.~Porto and J.~Pullin,
  ``Fundamental decoherence from quantum gravity: A pedagogical review,''
  arXiv:gr-qc/0603090.
  %%CITATION = GR-QC 0603090;%%
}

\Title{\vbox{\baselineskip12pt
%\hbox{\bf Draft}
%\hbox{\it Not for distribution}
\hbox{hep-th/0512200}
}}
{\vbox{\centerline{Observables in effective gravity}
}}
\centerline{{\ticp Steven B. Giddings}\footnote{$^\dagger$}
{Email address:
giddings@physics.ucsb.edu},  {\ticp Donald Marolf\footnote{$^*$}
{Email address:
marolf@physics.ucsb.edu}}, {\ticp James B. Hartle}\footnote{$^\#$}
{Email address:
hartle@physics.ucsb.edu}, }
\centerline{ \sl Department of Physics}
\centerline{\sl University of California}
\centerline{\sl Santa Barbara, CA 93106-9530}
%\bigskip
%\centerline{and}
%\centerline{{\ticp Other author}\footnote{$^\ddagger$}
%{Email address:
%} }
%\centerline{\sl wherever}
\bigskip
\centerline{\bf Abstract} We address the construction and
interpretation of diffeomorphism-invariant observables in a
low-energy effective theory of quantum gravity.  The observables
we consider are constructed as integrals over the space of
coordinates, in analogy to the construction of gauge-invariant
observables in Yang-Mills theory via traces.   As such, they are
explicitly non-local.  Nevertheless we describe how, in suitable
quantum states and in a suitable limit, the familiar physics of
local quantum field theory can be recovered from appropriate such
observables, which we term  `pseudo-local.' We consider
measurement of pseudo-local observables, and describe how such
measurements are limited by both quantum effects and gravitational
interactions.  These limitations support suggestions that
theFories of quantum gravity associated with finite regions of
spacetime contain far fewer degrees of freedom than do local field
theories.
%\draftmode
\Date{}

\newsec{Introduction}

An outstanding and central issue in the quantum mechanics of
gravity is the identification and interpretation of observables, see e.g. \refs{\Kuchar} and references therin. If gravity is studied about a background
with an asymptotic region, then this issue can be sidestepped, or at
least postponed, by focusing attention on the S-matrix and by
avoiding asking questions about local quantities within the
spacetime. Such backgrounds include the interesting cases of
asymptotically Minkowski and
asymptotically anti-De Sitter spacetimes\foot{For discussion of
the S-matrix in anti-De Sitter space, see \refs{\SBGBSM}. The status of an observable S-matrix in the De Sitter case is more controversial, but see \refs{\WittdS,\AstrodS}. }, but
not generic cosmologies. However, even in cases with an asymptotic
region, restricting attention to the S-matrix leaves out critical
physics; namely, the physics described by local observers  within the spacetime.
We are manifestly local observers within our own cosmological spacetime, and ultimately one of the goals of physics must be a precise mathematical description of the observations we make.

For many practical purposes, predictions can be made using the
formalism of quantum field theory in a curved background. However,
this puts aside the important problem of describing local physics
in a framework consistent with the expected symmetries and
properties of an effective low-energy quantum theory of gravity.
Moreover, we expect such a framework to be indispensable in any
attempt to describe the region near the singularity of a black
hole, the early universe, or more global aspects of quantum
cosmology.

In particular,  in field theory, the {\it local} observables of
the theory play a central role. However, the low-energy symmetries
of quantum gravity apparently include diffeomorphism invariance
which, as we will review, is known to preclude the existence of
local observables.  This leads to a well-known quandry in describing our own observations, which are
accomplished within the finite spacetime volume of the laboratory
or observatory. In particular, such observations take place on
time and distance scales that are quite small as compared to those
set by cosmology. It is clear that such observations are not
fundamentally described by a global S-matrix.

Thus, this paper works towards two important goals.  The first is
to improve our understanding of possible constructions of
diffeomorphism-invariant observables, which are the allowed
observables in quantum gravity.  The second is to find observables
that, in appropriate circumstances, approximately reduce to local
observables of field theory, or, more generally, to
``nearly-local" field theory observables, such as multilocal
expressions or Wilson loops.

As we will describe,
 in a wide class of circumstances, an approach to
the problem of defining diffeomorphism invariant observables is to
define quantities that are integrals (or multiple integrals) over
spacetime.  This is in rough analogy to using traces (or multiple
traces) to define gauge invariant observables in Yang-Mills
theory.

Given this, the next problem is to identify observables that, in an
appropriate sense, reduce to the local observables
of field theory.  A central idea here is that such observables
should be ``relational."  In classical general relativity, one
approach is to discuss the spacetime location of events relative
to some physical reference body, such as a clock on the earth.
Specifying events in this way allows one to build relational
classical observables, in an approach going back to \BryceI,
such
as the value of $R_{abcd}R^{abcd}$ at an event specified by its
relation to the earth and to the time registered on the clock.
Such quantities capture a certain sense of locality, but are
nevertheless {\it observables,} in the sense that they define
diffeomorphism-invariant functions on the space of classical
solutions. The question then is how, and to what extent, similar
correlations can be used to extract physical information in
quantum gravity.

The literature contains a number of approaches to this question,
see e.g.  \refs{\Kuchar,\BryceI\BryceII\Page\BanksTCP\TW\HartleCos\Smolin\Dolby-\GambiniYJ}.  The method of defining {\it
relational operators} has in particular been followed and extended
to the quantum context in
\refs{\BryceI,\BryceII,\TW,\Smolin,\RI\RII\RIII\ATU\QORD\BIX\MarT\Partial-\Pons}.    Here we
pursue this direction further,
 and argue  that this is the key to
 extracting physics that reduces to that of local field theory in
appropriate approximations.
These relational operators
are to be quantum analogues of the classical relational
observables discussed above.

Specifically, one of our main results will be to argue that, in
an appropriate limit, certain such relational, diffeomorphism
invariant observables of quantum gravity reduce to the more
familiar {\it local} observables of quantum field theory on a
fixed spacetime background.  We refer to such diffeomorphism-invariant observables as
``pseudo-local."
 An important point is that this reduction depends on the {\it
state} as well as the observable in question.

Our work represents
a field-theoretic generalization of similar results \refs{\QORD\BIX-\MarT} previously
established\foot{ In \refs{\MarT} such pseudo-local observables in 0+1 dimensional models were referred to
 as ``almost-local'' observables.} for certain
relational observables in various 0+1 dimensional systems
(reparametrization-invariant quantum mechanics).  {\bf These field-theoretic} observables suggest fundamental limits on locality and quantum measurement.  Moreover, growth in their fluctuations with the volume of space raises interesting questions in both the quantum cosmological and asymptotically flat contexts.

%A new feature is that the fluctuations of our observables grow with the volume of space.  Since one can arrange this growth to be modulated by an exponential suppression factor, this need not be problematic for cosmology.  Nevertheless, this feature raises interesting questions about connections between our pseudo-local observables and, for example,  the scattering observables of asymptotically flat spacetimes.  }

In outline, we first summarize the effective field theory approach
to gravity, describing its long-distance quantum dynamics and
symmetries,
as well as the problem of finding observables respecting
these symmetries.  In section three we investigate a broad class
of diffeomorphism-invariant quantities
that
 we expect to
serve as observables in gravitational physics in a manner similar to references \refs{\BryceI,\BryceII,\RI\RII\RIII\ATU\QORD\BIX\MarT-\Partial}.
Section four focuses on a special subclass of these observables,
the ``pseudo-local" observables, which in certain approximations
reduce to local observables of field theory; we do so primarily by
giving illustrative examples.
 Section five discusses
measurement theory of these observables.
Section six describes limitations on
observables arising from considerations of quantum mechanics and
gravity.  In particular, we see how general arguments (see e.g.
\refs{\BryceI,\CKN\GiLione\BanksWR-\GiLitwo})
concerning measurements in quantum gravity
manifest themselves in terms of restrictions on our relational operators.  Such limitations may
represent
 fundamental restrictions on
observation, and
 on the domain of validity of local quantum theories.   We close with a brief summary and discussion in section seven.

\newsec{Effective gravity, and the problem of observables}

As a fully controlled theory of quantum gravity does not yet
exist, we take an agnostic position here as to the nature of this
underlying fundamental theory; while, for
example, string theory could well be such a theory, as yet we lack
the ability to perform many calculations (particularly in the
non-perturbative regime).   However, whatever its dynamics, we
expect the fundamental theory to reduce to  quantum  general
relativity in non-planckian regimes.  Thus, our initial viewpoint
is that we will deal with the non-renormalizability of general
relativity by treating it as an effective theory with a cutoff at
$\roughly<\calo(\mpl)$, with a renormalization prescription
specifying the infinite number of couplings determined by the more
fundamental theory.  Ultimately, we will find further constraints
that suggest the need to supplement this cutoff with more
stringent limitations on the effective theory.

While we will not be precise about the nature of the cutoff, in
our view a central question is in what regime the low-energy
effective theory predicts its own failure; before this one expects
that the precise cutoff prescription has negligible effect, and
beyond this we will need the full quantum dynamics of the
underlying theory to make predictions.

Although we do not know the fundamental description of states in
quantum gravity, we expect that the cutoff theory has an effective
description similar to that obtained by canonical quantization of
the gravitational field.  In particular,
there should be a regime in which states $|\Psi \rangle$ admit an effective
description in terms of functionals  $\Psi[h_{ij}, \phi^r]$ of a
Euclidean signature three-metric (or in greater generality
a $D-1$-metric) and other fields $\phi^r$ on some  surface $\Sigma$,
where in the classical limit $\Sigma$ will become a spacelike three-surface
embedded in some four-dimensional spacetime.

The canonical formalism provides a
useful perspective on the long distance quantum dynamics of
gravity.  In addition to any symmetries of the matter theory, the low-energy
symmetries of the theory should include diffeomorphisms,
$x^\mu\rightarrow x^\mu + \xi^\mu(x^\nu)$.
As a consequence, we learn from this
formalism (see, e.g., \refs{\Dirac,\BryceII}) that this dynamics
should be described by a set of constraints of the form
\eqn\wdw{\calh |\Psi\rangle =0, \ \ \ \calh_i |\Psi\rangle =0,}
where $\calh$ is the densitized scalar constraint (sometimes called the ``Wheeler-DeWitt operator''),
\eqn\hwdw{\calh = G_{ijkl}   \pi^{ij} \pi^{kl}  - \sqrt{h}\left[ {}^3 R(h) + {16\pi\over \mpl^2} \calh^m\left( \pi_r, \phi^r,h\right) \right]\ ,}
and $\calh_i$ are the densitized vector constraints
\eqn\Hidef{\calh_i = {{16 \pi} \over {M_p^2}} \left(- 2D_j \pi^{ij} \ + \calh_i^m \right) .}%
Here $D_i$ is the covariant derivative on $\Sigma$ compatible with $h_{ij}$.
In the above, the superspace metric $G_{ijkl}$ is
\eqn\gsuper{G_{ijkl} = \hf \left( {{16 \pi} \over {M_p}} \right)^2 {1 \over \sqrt{h}} (h_{ik} h_{jl} + h_{il} h_{jk} - h_{ij} h_{kl} ),}
while $\calh^m, \calh^m_i$ represent contributions from the
matter fields, and $\pi^{ij}, \pi_r$ are the momenta conjugate to
$h_{ij}, \phi^r$.  In particular, in the wavefunctional
representation described above,  $\pi^{ij}, \pi_r$,  will act as
$-i {\delta \over {\delta h_{ij}}}$, $-i {\delta \over {\delta
\phi^r}}$.   (In greater generality, initial conditions or
processes that cause the Universe to branch may, in a third
quantized framework\refs{\BanksJE,\GiddingsWV} introduce non-zero terms on the
right hand side of \wdw; for further discussion see
\refs{\GiddingsWV}.)  Proper definition of these operators requires
an appropriate operator ordering and regularization, which we view
as being supplied by our cutoff prescription.

 Although the constraints $\calh,\calh_i$ encode invariance under
diffeomorphisms, they generate a somewhat different algebra known
as the ``hyper-surface deformation algebra,''

\eqn\HypDefA{\eqalign{
\left[
\int_\Sigma N \calh, \int_\Sigma M \calh
\right]
 &= i{{16 \pi} \over {M_p^2}}  \int_\Sigma (N \partial_i M - M \partial_i N) h^{ij} \calh_j\ , \cr
  \left[ \int_\Sigma N^i \calh_i, \int_\Sigma M^j
\calh_j \right]&= i {{16 \pi} \over {M_p^2}} \int_\Sigma [\vec N, \vec M]^k  \calh_k\ ,\cr
 \left[ \int_\Sigma N \calh, \int_\Sigma M^j \calh_j
\right] &= - i {{16 \pi} \over {M_p^2}} \int_\Sigma  {\cal L}_{\vec M} N  \calh\ , } }
where ${\cal L}_{\vec M}$ denotes the Lie derivative along the
vector field $M^j$ and $[\vec N, \vec M]$ denotes the commutator
of the two vector fields.  The operators $\int_\Sigma N^i \calh_i$
generate diffeomorphisms of $\Sigma$ and, on classical solutions,
the operators $\int_\Sigma N \calh$ generate displacements of the
hypersurface $\Sigma$ along the vector field $N n^\mu$, where
$n^\mu$ is the future-pointing spacetime normal to $\Sigma$.  As a
result, the hyper-surface deformation algebra generates the same
orbits in the space of classical solutions as does the
diffeomorphism group \refs{\IKI,\IKII}; i.e., invariance of a
function on the space of solutions under the action of one algebra
is equivalent to invariance under the action of the other algebra.
Similarly, invariance under the constraints
$\calh, \calh_i$ should also encode diffeomorphism invariance in the low
energy effective description of quantum gravity.

The lack of local observables in gravity is now clear. As first
emphasized by Dirac \refs{\Dirac}, a predictive framework requires
that observables commute with the generators of gauge symmetries.
However, for example, given any local scalar field $\phi(x)$, this
field commutes with the constraints if and only if it is invariant
under all diffeomorphisms; i.e., if $\partial_\mu \phi$ vanishes
identically.  Similar results follow for spinor, vector, and
tensor fields.  Hence, local fields are not observables in
theories with gravity.

Now, one could take the viewpoint that we cannot even
approximately identify gauge-invariant observables until we have
total control over the fundamental theory of quantum gravity.
However, this seems an extreme position if there is a sensible
cutoff theory of effective gravity at low energies. The reason is that observables
should exist in the effective theory, and such observables should
respect the low-energy gauge invariance.  Put differently, we believe that we should be
able to describe the low energy observations of local observers
in terms of the framework of low energy gravity. While an exact
identification of the observables of quantum gravity presumably
requires the ultimate fundamental theory of quantum gravity,  we
expect that a framework for treating them in the low-energy theory
will remain useful.

\newsec{Diffeomorphism-invariant observables}

As reviewed above, the problem of finding quantum gravity
observables is that of finding the appropriate
gauge-invariant operators.
Moreover, the ones capable of describing our
experiences in the laboratory should reduce to the usual local
observables of quantum field theory in an appropriate limit.

Beginning with the first question, in efffective gravity, we seek
operators that are combinations of the metric and other fields
$\phi^r$, which are hermitian,\foot{As discussed below, we will
use the {\it induced} or {\it group averaging} inner product
\refs{\QORD,\Higuchi\Landsman\ALMMT\GMII-\OS} on the space of physical states (i.e.,
those satisfying \wdw), so that operators which are hermitian with
respect to the inner product on the auxiliary Hilbert space are
automatically hermitian on the physical Hilbert space. However, due to the complicated nature
of the operators we consider, self-adjointness can be more subtle.  See \refs{\ToF} for comments on this issue and an example of how it may be dealt with.} and
which commute with the constraints, $\calh$, $\calh_i$. For
example, let ${\hat \calo}(x)$ be a local scalar observable in
ordinary quantum field theory; in a scalar theory, we might
consider $\cohat(x) = \phi(x),\, \phi^2(x),\, \ldots$.  Such an
operator is not diffeomorphism invariant, but
\eqn\difop{\calo = \int d^4 x \rtg \cohat(x)}
is clearly diffeomorphism invariant.  It also commutes with the
constraints ${\cal H}, {\cal H}_i$. The key step in this argument
is that we define the time-dependence of $\hat {\cal O}(x)$ in
\difop\ through the Heisenberg equation of motion \eqn\Hei{ i
{\partial \over {\partial t}}  \hat \calo (x) = [\hat \calo (x),
\int_\Sigma (N {\cal H} + N^i {\cal H}_i ) ]. } Thus, the
analogous commutator with $\calo$ reduces directly to a boundary
term, which vanishes under appropriate boundary
conditions\foot{More discussion of boundary conditions will
follow.  For examples in the 0+1 context, see
\refs{\QORD,\MarT,\GMII,\WhereAreWe}.  In particular, convergence
of the integral in \difop\ (and in (3.3) below) is a subtle issue:
The integral converges on what is called the auxilliary Hilbert
space below, but this space may contain no normalizable states
satisfying the constraints \wdw.  Nevertheless, the action of
$\calo$ on this auxiliary Hilbert space induces an action on
physical states.}.   It is also clear that \difop\ is invariant
under spatial diffeomorphisms, and therefore that it commutes with
any operator of the form $\int_\Sigma \tilde N {\cal H}$ where
$\tilde N$ is related to $N$ in \Hei\ by a spatial diffeomorphism.
We may combine these observations to show that $\calo$ commutes
with ${\cal H}, {\cal H}_i$. The corresponding fact is explicitly
shown in a number of 0+1 models in \refs{\QORD}, which paid close
attention to subtleties such as the implicit appearance of $N,N^i$
in \difop\ (through the time-dependence of $\hat {\cal O}(x)$).

More generally, for a collection of matter fields $\phi^r$,
consider an arbitrary local scalar  density formed from the
the fields, the metric, and their derivatives which is invariant
under any gauge symmmetries of the matter theory, %%
\eqn\ohatex{\cotilde = F(\phi^r(x),\partial_\mu\phi^r(x),\ldots;
g_{\mu\nu}(x), \partial_\lambda g_{\mu\nu}(x), \ldots)\ .} Then
\eqn\difopII{\calo = \int d^4 x \ \cotilde (x)}
will commute with
the constraints $\calh$, $\calh_i$ and is an observable if
$\cotilde$ is hermitian. We refer to observables of the form
\difopII\ as ``single-integral observables.'' Clearly we can
formulate other operators that are likewise diffeomorphism
invariant, but which are more complex, by considering operators that depend
on more than one point.  Examples would be objects such as
\eqn\nonlocex{\calo = \int d^4x\rtg\int d^4 y \rtg f(\phi(x),\phi(y))\ , }
generalizations of Wilson loops, and other such ``multilocal"
expressions.

To describe local experiments, we
will be interested in
 such
 observables which (approximately)
localize in some spacetime region, and the corresponding operators
will need to include
physical degrees of freedom which specify this
region. This connects to
a perspective going back to Einstein \refs{\EinsDov}, and
emphasized by DeWitt\refs{\BryceI,\BryceII}, which we may paraphrase as follows:
the description of the flow of time requires a
self-consistent inclusion of the actual dynamical degrees of
freedom that register this flow. We follow an established
tradition and refer to such degrees of freedom as a {\it clock},
though we emphasize that the reading of this clock need not be
simply related to the passage of proper time as defined by some
metric, and though more generally we are interested in position
information in both space and time directions.  We hope that this
terminology does not cause excessive confusion.

In preparation for proceeding, let us make three comments. First,
we will be most interested in operators $\cotilde$ which are
composite, and such operators require a regularization in order to
be defined in quantum field theory.  We assume this is provided by
the cutoff of the effective gravity theory, and that appropriate
renormalization prescriptions are provided at that cutoff scale.
Secondly, note that single integral observables are precisely the
operators that can be added to the action to give a local
interaction term in the low-energy effective gravity theory.
Finally, while the integrals in {\it e.g.} \difop, \difopII,
\nonlocex\ formally may extend into regions where the effective
gravity description begins to fail, we assume that there are
appropriate operators and states for which the contribution of such regimes
is small.  We will elaborate more on this point subsequently.

Before considering details of the problem of localization, we
finish this section by discussing the formal role of
diffeomorphism-invaraint observables in a theory of gravity. In
particular, we will discuss details of defining matrix elements of
 the above diffeomorphism-invariant observables between physical states;
i.e., between states satisfying the constraints \wdw.  (We will
discuss the relation of such matrix elements to measurements in
section five.) This discussion is rather technical.  The reader may wish to scan the rest of this section quickly on a first reading of the paper.

Computation of matrix elements requires an
inner product on the
space of physical states.  Here we follow an approach described
in \refs{\QORD,\Higuchi\Landsman\ALMMT-\GMII\OS,\WhereAreWe,\MarOne}
which define the  {\it induced} or {\it group averaging} inner
product on physical states\foot{See \refs{\WhereAreWe} for a brief introduction to the method and \refs{\HarMar} for comments on how, in a mini-superspace context, the positive definite induced inner product can be related to the more familiar Klein-Gordon inner product.}.  This inner product also agrees with
certain BRST methods \refs{\OSII}.

As a first step, we may note that the space of functionals of the
metric and fields ({\it i.e.} not necessarily satisfying the
constraints) can be made into a Hilbert space via the
usual Schr\"odinger representation inner product\foot{In fact, in
field theory the particular inner product used needs to be adapted
to the dynamics of the theory. However, at the formal level at
which we work here, all such details are taken care of by the path
integral and the renormalization process.}.  However, in general
{\it no} states satisfying the constraints \wdw\ will be
normalizable in this inner product.
The physical inner product can at
best be a ``renormalized'' version of the auxiliary inner
product.\foot{Some constraints may have both normalizable and
non-normalizable solutions, in which case one expects that these
two classes define different superselection sectors.  One expects
similar superslection rules between classes of states whose norms
in the auxiliary space in some sense have different degrees of
divergence.  See \refs{\ALMMT,\GMII, \MarOne, \AD\JLI-\JLII}.}
For this reason, we follow
the tradition of referring to the resulting
Hilbert space as the {\it auxiliary} Hilbert space.  We denote the
corresponding (auxiliary) inner product as $\langle \Psi_2 \pip
\Psi_1 \rangle$. States in the auxiliary Hilbert space may be
expanded, for example, in terms of the
 basis of eigenstates
 $\pip{h_{ij}, \phi^r} \rangle$
of the configuration variables\foot{In fact, due to our desire to
perform the integrals \difop,\difopII, we work in a Heisenberg
picture in which the operators  $h_{ij}, \phi^r$ depend on time.
By $\pip{h_{ij}, \phi^r} \rangle$, we mean the eigenstate of
$h_{ij}, \phi^r$ on some (fixed but arbitrary) reference
hypersurface $\Sigma_0$ in the space of coordinates $x$.}
 $h_{ij}, \phi^r$.

We may usefully combine the step of solving the constraints with the step of introducing a useful inner product on the space of solutions.
  In particular, consider the functional integral
\eqn\prodstat{ \langle {h_2,\phi_2^r} \pip \eta \pip
{h_1,\phi_1^r}\rangle := \int_{h_1, \phi^r_1}^{h_2,\phi^r_2} \cald
g \cald \phi^r e^{iS}\ ,}
where we have taken this integral to define the matrix elements of
an object $\eta$.\foot{If the inner product on the auxiliary
Hilbert space was chosen appropriately, \prodstat\ and linearity
should at least define matrix elements $\langle \Psi_1 \pip \eta
\pip \Psi_2 \rangle$ of $\eta$ when $\pip\Psi_1 \rangle,
\pip\Psi_2 \rangle$ lie in a dense subspace $\Phi$ of this Hilbert
space, though $\eta \pip \Psi_2 \rangle $ itself may not be a
normalizable state. Instead, the image of $\eta$ naturally
consists of linear functionals on $\Phi$, which is sufficient for
our purposes.  See \refs{\MPI} for a discussion of the path
integral, and \refs{\Landsman\ALMMT-\GMII,\WhereAreWe,\MarOne} for
a more general discussion of this point.  See also
\refs{\SumOver,\Projector} in the context of loop quantum
gravity.} Here $S$ is the action, $h_i, \phi_i^r$ specify data on
initial and final slices, and we functionally integrate over all
interpolating geometries and field configurations, with an
appropriate gauge-fixing procedure. While we have written
\prodstat\ in a covariant notation, the functional integral we
have in mind is most easily defined using the canonical form of
the functional integral in which $S = \int d t d^3x \left(N \calh
+ N^i \calh_i\right)$ where $N,N^i$ are the lapse and shift.

In particular, we take the
integral $\cald g$ above to include an integral over both positive
and negative lapse.  An important consequence of this is that,
as noted in e.g. \refs{\HarHal}, the
functional integral \prodstat\ satisfies the constraint equations
\wdw\ in {\it both} arguments.  That is, we have
\eqn\solve{ \langle {h_2,\phi_2^r} \pip \calh \eta \pip   {h_1,\phi_1^r}\rangle =  \langle {h_2,\phi_2^r} \pip \eta \calh \pip   {h_1,\phi_1^r}\rangle =0,}
and similarly for $\calh_i$. The operator $\eta$ is often called a
``rigging map;''
roughly speaking, we may think of $\eta$ as a functional delta
function $\eta\sim \Delta[\calh,\calh_i]$ which enforces the entire
set of constraints. We see that the image of
 $\eta$ consists of solutions to the constraints and, in
addition, we see that any state of the form $\calh \pip\Psi
\rangle$ is annihilated by $\eta$.  Thus $\eta$ is highly
degenerate, and we may think of $\eta$ as identifying entire
equivalence classes, denoted $|\Psi \rangle$, of auxiliary states $ \pip\Psi
\rangle$ with solutions of the constraints.   Thus, we may think
of the equivalence classes $| \Psi \rangle$ as physical states
themselves; i.e., $|\Psi \rangle = \eta \pip \Psi \rangle$.  Note that the projection $|h,\phi^r\rangle = \eta \pip h,\phi^r\rangle$ results in an {\it overcomplete} basis of physical states.

The integral over both positive and negative lapse in \prodstat\ also implies
$\eta$ is hermitian.  It thus defines an inner product, which we shall denote in the usual Dirac fashion, on the equivalence classes
$| \Psi \rangle$:
\eqn\phys{ \langle \Psi_1|\Psi_2 \rangle : = \langle \Psi_1 \pip\eta \pip \Psi_2 \rangle.}
As discussed in \refs{\MPI}, the inner product \phys\ defined in
this way by \prodstat\ agrees with what is known as the {\it
induced} (or group averaging) inner product. If \phys\ is positive
definite,\foot{See \refs{\MarOne,\GMII,\WhereAreWe,\JLIII}  for
known results concerning this positivity.}
it defines a
Hilbert space of physical states.

 Given that $[\calo,\calh]=[\calo,\calh_i]=0$, the observable
$\calo$ preserves the space of physical states\foot{In fact, because $\eta$ is built from $\calh$ and $\calh_i$, $\calo$ commutes with $\eta$ in the
sense that $\calo \eta \pip \Psi \rangle = \eta \calo \pip \Psi
\rangle$.}; i.e.,
\eqn\checkO{\calh \calo |\Psi \rangle = \calh_i \calo |\Psi
\rangle =0.} As a result, the above definitions allow us to
compute the matrix element of an observable $\calo$ between two
physical states; we can act with $\calo$ on state $|\Psi_1 \rangle = \eta \pip
\Psi_1\rangle$, and then take its induced product with $|\Psi_2 \rangle = \eta \pip
\Psi_2\rangle$, in the usual fashion:
\eqn\matdef{ \langle \Psi_2| \calo  | \Psi_1\rangle\  := \langle
\Psi_2\pip  \calo\eta \pip \Psi_1\rangle\ .}
Note that since \matdef\  is the physical inner product of $\calo
\eta \pip \Psi_1 \rangle$ and $\eta \pip \Psi_2 \rangle$, the
result depends only on the choice of physical states $|\Psi_1
\rangle, |\Psi_2 \rangle$ and not on the particular
representatives $\pip \Psi_1 \rangle, \pip \Psi_2 \rangle$ of the
corresponding equivalence classes.

So far we have outlined the definition of a rather broad class
of operators which are manifestly non-local. As yet, we have made
no direct contact with local observables in quantum field theory.
However, in the sections below we explore how, in an appropriate
approximation, certain diffeomorphism-invariant operators do
indeed reduce to the local observables of ordinary quantum field
theory, with one critical caveat:  such a reduction depends
essentially on the choice of state $|\Psi \rangle$ in combination
with the choice of observable $\calo$. These points are best
illustrated by examples, to which we now turn.

\newsec{Diffeomorphism invariant observables and localization: examples}

In the last section we outlined the general low-energy effective
framework for quantum gravity, emphasizing that observables are
necessarily invariant under the constraints, and that such operators are naturally given by diffeomorphism-invariant expressions such as \difop, \nonlocex.  As noted above,
diffeomorphism invariant operators are not local, so that
additional steps are required to mesh this discussion with our
usual treatment of local physics.  We attempt to fill
 this gap here through a treatment of a number of examples.

Before beginning, let us recall from the last section that a
critical step is to define diffeomorphism-invariant observables on
the {\it auxiliary} Hilbert space.  If this can be done, then such
operators naturally define observables on the physical Hilbert
space as well.  As a result, we may reasonably hope to separate
the treatment of some issues of locality from a detailed study
of, say, the constraints \wdw.  For this reason, we
begin our first two examples by working with
diffeomorphism-invariant operators in the context of scalar field
theory in the usual (unconstrained) Fock space, before considering
coupling to the 3+1 gravitational field.
In contrast, our last
two examples will directly include the gravitational field, albeit
in low dimensions (0+1 and 1+1) where the dynamics of gravity is
somewhat trivial.

\subsec{Scalar fields as physical coordinates: the $Z$ model}

As our first example, we discuss pseudo-local
diffeomorphsim-invariant observables constructed using scalar quantum field
theory.  This may be regarded either as a toy model that
illustrates some features of interest, or as a first step toward
studying
pseduo-local observables in low-energy gravity coupled to a set
of such scalars. In particular, we will see how one can,
approximately and in an appropriate state, connect
pseudo-local observables to the usual framework of local
observables of quantum field theory. In doing so, the key point is
that the location of the local observable is specified {\it
relative} to a structure determined by the state in a manner
determined by the particular pseudo-local observable. We believe
that this example serves as a paradigm for how the local operators
of field theory can be recovered in theories with diverse field
content.

Our starting point is a general theory with fields $\phi^a$. We
work in four-dimensions, although the na\"\i ve generalization to
higher dimensions follows trivially; we initially consider working
in a flat background, but discuss aspects of curved spacetimes
shortly. To define the $Z$-model corresponding to the field
theory, we assume that in addition to the fields $\phi^a$ we have
four additional massless free scalar fields $Z^i$, $i=0,1,2,3$.
For such a theory, we may consider an initial state
$|\Psi_Z\rangle$ such that, in some spacetime region of spacetime,
\eqn\Zvev{\langle\Psi_Z |Z^i|\Psi_Z\rangle = \lambda \delta^i_\mu x^\mu\ ,}
that is, the fields have
expectation values
that satisfy the classical equations of motion
and moreover are proportional to the background coordinates.  The
state of these fields therefore spontaneously breaks the
Poincar\'e invariance of the background spacetime.  In particular,
we will take $\Psi_Z$ to be a minimally excited such state, in the sense that we take the fluctuating field
\eqn\fluctdef{{\tilde Z}^i = Z^i - Z^i_{cl} = Z^i- \lambda \delta^i_\mu x^\mu}
to be in the Fock
ground state.

The basic idea is that positions of local observables can be defined in a translation invariant way relative to the background
expectation values \Zvev.  Specifically, given a local operator
$O(x)$ in the theory of the $\phi^a$'s, we might imagine defining
operators of the form
\eqn\naopdef{\calo_{0,\xi} = \int d^4 x O(x)
\delta[Z^i(x) - \xi^i] |{\partial Z^i\over \partial x^\mu}|\ .}
Such operators were suggested in \refs{\BryceI},
though we will treat them directly in quantum field theory without
first passing to the semi-classical limit. For a classical
solution of the form \Zvev, the delta function picks out a
definite point.
Moreover, it will pick out a finite set of points in a generic
perturbation of \Zvev.  Thus, operators of the form \naopdef\
qualify as pseudo-local observables.

The operator defined in \naopdef\ is not only Poincare invariant,
but also diffeomorphism invariant under changes of coordinates
$x^\mu\rightarrow x^{\mu \prime}(x^\nu)$.
$\calo_{0,\xi}$ is, however, potentially problematic to define
in the context of a quantum field theory due to the
$\delta$-function of quantum fields in \naopdef. For this reason,
we instead consider a similar but more regular operator of the
form
\eqn\opdef{ \calo_{\xi} = \int d^4 x O(x) e^{-{1\over\sigma^2} (Z^i-\xi^i)^2} |{\partial Z^i\over \partial x^\mu}|\ ,}
where $\sigma$ is a constant of mass dimension one that plays the
role of a {\it resolution} of the
operator in
 \naopdef.

Suppose now that
we evaluate the expectation value of a product
of a collection of $N$ such operators, each with different
$\xi_A^i$, $A=1,\ldots,N$, in a state of the form
\Zvev. We might expect that this expectation value approximately
reduces to the correlation function of  a product of the operators
$O(x_A^\mu)$, with locations given by
\eqn\xlocate{x^\mu_A = {1\over \lambda} \delta^\mu_i \xi^i_A\ .}
Let us examine this calculation more closely in order to check
this statement, and also to find its limitations.

The functional integral computes the correlation function, in the state $|\Psi_Z\rangle$, time-ordered with respect to parameter time,
\eqn\corrdef{\langle T( \calo_{\xi_1}\cdots \calo_{\xi_N})\rangle
= \int\cald \phi^a \int _{\Psi_Z} \cald Z e^{i S[\phi^a]+ i
S[Z]}\prod_A^N \calo_{\xi_A}\ .}
Here, as we've indicated, the boundary conditions
on the $Z$ integral are furnished by the state giving \Zvev.  We
assume that the gaussian operators in $Z$ are determined in some
regularization scheme, by a set of operator boundary conditions,
which we assume preserves the correct semiclassical limit for the
gaussian. A convenient way to evaluate this expression is to
Fourier transform,
\eqn\ftgauss{e^{-{1\over\sigma^2} (Z^i-\xi^i)^2} = {  {\sigma^4} \over {16 \pi^2}}
\int d^4\kappa e^{ - {{\kappa^{2}\sigma^2} \over 4} +i\kappa_i(Z^i-\xi^i)}\ .}
We then write $Z^i$ as a classical piece plus fluctuation piece,
as in \fluctdef, %%
%\eqn\zexp{Z=Z_{cl}+{\tilde Z}}
%
 and functionally integrate over $\tilde Z^i$ to find
\eqn\zcorr{\eqalign{\int _{\Psi_Z} &\cald Z e^{iS[Z]} \prod_A^N e^{-{1\over\sigma^2} (Z^i(x_A)-\xi^i)^2}|{\partial Z^i\over \partial x^\mu}| = \int \prod_{A}(
{  {\sigma^4} \over {16 \pi^2}}
 d^4\kappa_{A} ) e^{iS[Z_{cl}]} \cr &
\left[e^{-\sum_A \kappa_A^2\sigma^2/4 +i\kappa_{A,i} (\lambda x_A^i -\xi_A^i)} e^{-{i\over 2} \sum_{AB} \kappa_A\cdot \kappa_B G(x_A,x_B)} \left|{\partial Z^i_{cl}\over \partial x^\mu} \right|  M(x_A, \kappa_{A,i})\right]\ .}}
Here $G(x_A,x_B)$ is the appropriate Green's function and $M$ is a factor arising from the fluctuation part of the jacobian.  The first exponent is the classical action for $Z_{cl}$, the second is the contribution of the classical solution to the correlation function, and the third comes from fluctuations of the fields $Z$ about $Z_{cl}$.
The correlation function \corrdef\ then incorporates this expression as
\eqn\totcorr{\int\prod_A dx_A \langle T(O(x_1)\cdots
O(x_N))\rangle_\phi \int _{\Psi_Z}  e^{iS[Z]}  \cald Z \prod_A^N
e^{-{1\over\sigma^2} (Z^i(x_A)-\xi^i)^2} \left|{\partial Z^i\over
\partial x^\mu}\right|\ ,}
where the notation $\langle \cdots \rangle_\phi$ denotes a correlator in the vacuum of the $\phi$ theory.

If we can neglect the fluctuation pieces of \zcorr, this expression reduces to the usual field-theory correlator of the $O(x_A)$'s, smeared over a width
\eqn\xgauss{
\Delta x_A\sim {\sigma/\lambda}}
about the values \xlocate,
 \eqn\opapprox{\langle T(\calo_{\xi_1}\cdots
\calo_{\xi_N})\rangle \approx \int\cald \phi^a
e^{iS[\phi^a]}[O(x_1)\cdots O(x_N)]\ .} The operator products
$\calo_{\xi_1}\cdots \calo_{\xi_N}$ (without time-ordering)
behave similarly.

Fluctuations correct this expression.  One can estimate their
sizes by expanding $Z$ as in \fluctdef\ and extracting the leading
(quadratic) term.  Equivalently, without the jacobian factor
$\left|{{\partial Z} \over {\partial x}}\right|$, the requirement
that they be small follows immediately from the form of \zcorr,
\eqn\correstone{{1\over \sigma^2} \langle T\left( {\tilde Z^i}(x_A) {\tilde Z^j}(x_B) \right) \rangle = {{\delta^{ij}}\over \sigma^2} G(x_A,x_B)\sim {{\delta^{ij}}\over \sigma^2}  {1\over (x_A-x_B)^2}\ll 1\ .}
Including contributions of the jacobian, we also find the conditions
\eqn\corresttwo{\eqalign{{1\over \sigma\lambda} \langle T\left(\partial {\tilde Z^i}(x_A) {\tilde Z^j}(x_B) \right) \rangle & \sim {{\delta^{ij}} \over \sigma^2}  {1\over (x_A-x_B)^2} {\sigma\over \lambda} {1\over(x_A-x_B)}  \ll1\cr
{1\over \lambda^2} \langle T\left( \partial {\tilde Z}(x_A) \partial {\tilde Z}(x_B) \right) \rangle &\sim {{\delta^{ij}}\over \lambda^2} {1\over (x_A-x_B)^4}\ll1\ .}}

One can begin to understand these conditions by considering
working in an effective theory\foot{However, it is interesting to
note that, even for our highly composite operators \opdef, the
correlators of operator products (without time-ordering) are
well-defined and approximate correlators of $\phi(x_1)\cdots
\phi(x_N)$ without any such cut-off, so long as the theory of the
$\phi$-fields is itself well-defined. In particular, $\calo_\xi$
is a densely defined operator on our Fock space. These results will be presented in a forthcoming paper.}
 with a momentum
cutoff $\Lambda$. In such a theory, there is effectively a bound
\eqn\xbound{{1\over |x_A-x_B|}\roughly< \Lambda\ .}
Saturating this bound gives the tightest constraint from \correstone:  $\sigma\gg\Lambda$.  The gaussian uncertainty \xgauss\ in the positions $x_A$ is determined by $\sigma$ and $\lambda$.
The field momentum $\lambda$ should be bounded by $\Lambda^2$, for validity of the cutoff theory.  These statements then translate into a lower bound on the uncertainty in $x_A$:
\eqn\xunc{\Delta x\gg {1\over \Lambda}\ .}
This result is sensible: in the context of the cutoff theory, the maximum distance resolution is the inverse of the cutoff. These results are readily generalized to other dimensions.

The constraints \correstone, \corresttwo\ are due to basic
quantum uncertainty in the definition of the position using the
relation to the state of the $Z$ fields. While they have been
derived directly only in this model, we expect
similar results to hold for an arbitrary model in which the
location at which an observable is being computed is
determined by a physical dynamical clock or position variable localized in the
region being investigated.\foot{However, so long as the region studied is not the entire universe, we
leave open the possibility that pseudo-local
observables may exist for which the clock and position degrees of freedom are kept at some distance from the region under investigation.  Such ``remote sensing'' observables are particularly relevant to spacetimes with an asymptotic region.}  The reason is that they follow simply
from the uncertainty principle and from the properties of a
theory with a cutoff.

We
now complete our discussion of the $Z$ model  by making a few
comments on the generalization to
include a dynamical metric.  Note that both the $Z$
fields and the $\phi^a$'s
will couple to the metric.  We can
 consider a combined state of the $Z$ field and metric such that
the behavior of the $Z$ fields is approximately classical;
the
weakest version of this is simply that the expectation values of
the $Z$'s vary monotonically, and that their fluctuations are
small.  In this case, the $Z$'s approximately define temporal and
spatial location,
in a manner analogous to the above discussion.  In fact, in such
a case, the operators \opdef\ are already diffeomorphism
invariant.  In some cases one might also want to consider
a similar but different set of diffeomorphism invariant operators,
\eqn\gopdef{ \calo_{g,\xi} = \int d^4 x \sqrt{-g} O(x)
e^{-{1\over\sigma^2} (Z^i-\xi^i)^2} \ ,}
where the determinant in \opdef\ has been replaced by
$\sqrt{-g}$.  Such observables also
approximately localize,
subject to constraints analogous to \correstone, \corresttwo.

%However, in the case of dynamical geometry, one does not expect
%\Zvev\ to provide a viable classical background over an
%arbitrarily large region. In particular, the constant energy
%density will back-react on the geometry, but nonetheless we expect
%to be able to use such operators to determine position within a
%sufficiently small region. We will further discuss constraints
%that arise from the incorporation of gravity in section six.
However, in the case of dynamical geometry, one does not expect
\Zvev\ to provide a viable classical background over an
arbitrarily large region. In particular, the constant energy
density will back-react on the geometry.  It is therefore natural to consider states in which the $Z$-fields approximate \Zvev\ only over some region $\Omega$ which is bounded in space (though which need not be bounded in time if the physics provides a way to keep the $Z$-fields from dispersing).  The $Z$-fields would then be essentially in their vacuum state outside of $\Omega$.    In this context, we say that only the region $\Omega$ has been `instrumented' with our dynamical  reference background.

One expects
to be able to use the operators \opdef\
to determine position within  the region $\Omega$.
We will further discuss constraints
that arise from the incorporation of gravity in section six, but one effect which must now be taken into account arises directly from the scalar sector in the region $\Omega^c$ which forms the complement of $\Omega$; i.e., from the region {\it outside} of the original region $\Omega$.  The effect of this region can be modeled by simply computing correlators of ${\cal O}_{\xi}$ in the vacuum state $|0 \rangle$.  One can easily arrange that, in $|0\rangle$, the integrand of \opdef\ has vanishing expectation value,  by shifting the operator.  Thus $\Omega^c$ does not contribute to the expectation value of ${\cal O}_{\xi}$.  Nevertheless, it will  in general contribute to the expectation value of ${\cal O}_{\xi_1}{\cal O}_{\xi_2}$; i.e., to correlators of pseudo-local observables, and thus to the fluctuations of pseudo-local observables about their expectation values.

When considering a fixed observable ${\cal O}_{\xi}$, it is clear that  for sufficiently large $\Omega^c$  the resulting noise will overwhelm our desired signal. In particular, our signal will be overwhelmed in an infinite volume universe.  When the volume of space is merely very large (but finite), this effect will place fundamental limits on the accuracy with which any given ${\cal O}_{\xi}$  reduces to a local observable in a given region.  However, since the fluctuations involve the operator $e^{-{1\over\sigma^2} (Z^i-\xi^i)^2}$, they are exponentially small in the parameter $(\xi^i)^2$. Thus, such limits need not be especially stringent in practice and can be further suppressed by using operators that effectively enforce more conditions.  On the other hand, they raise interesting questions concerning the infinite volume limit and the connection to, for example, the S-matrix. They may also play an interesting role for universes which experience sufficiently long periods of rapid growth, and in particular in eternal inflation scenarios.

\subsubsec{Generalizations}

An important overall goal of this work is to understand some
approximation of the types of observations we make, for example,
at particle accelerators such as the LHC.  The $Z$ model captures
some aspects of such observations, in particular their
localization, but in reality experimental apparatuses are quite
complex and involve detectors which are very complicated excited
states above the vacuum.  Working towards actual physical
measurements, one may wish to consider more complicated operators
than those in \opdef, \gopdef.  One first step is to separate the
{\it timing} function from the {\it observing} function, for
example by considering both the $Z$ fields and additional degrees
of freedom comprising a detector.  A candidate class of
diffeomorphism-invariant observables is of the form
\eqn\clmeas{ \calo_{g,\xi} = \int d^4 x \sqrt{-g} O(x) m(x)
e^{-{1\over\sigma^2} (Z^i-\xi^i)^2}\ .}
Here $m(x)$ is an operator acting on the detector.  Concretely, $O(x)$ might be an operator annihilating a photon, with $m(x)$ describing the consequent excitation of an atom (or more complicated ensemble).  One might choose the $Z$ operators to merely provide approximate location information, which for example could be much less accurate than the time scale associated with the spacing between the detector's energy levels.  Clearly there are further extensions of increasing complexity.

\subsec{$\psi^2\phi$ model}

We next consider another field theory example which illustrates some of the
 features of pseudo-local observables.
  Specifically,
consider a
 theory of two massive non-interacting scalar fields, $\psi$ and
$\phi$. In this case, an example of a generalized observable is
the diffeomorphism-invariant operator
\eqn\psf{\calo_{\psi^2\phi} = \int d^4 x \sqrt{-g} \psi^2(x)
\phi(x)\ ,}
which has the virtue of being simpler than the gaussians of the $Z$ model, as well as renormalizable.

Despite the simplicity of such operators, localized information
about $\phi$ can be obtained by encoding this information in the
state of the $\psi$-field. This is a second paradigm for recovery
of local operators from diffeomorphism-invariant operators. For
example, begin by working about a flat background, and suppose
that we are interested in extracting an $N$-point function of the
field $\phi$ from a correlation function of the operators \psf. We
do so by considering $\psi$ states corresponding to incoming and
outgoing wavepackets.  These are defined in terms of wavepacket
creation operators, which, for a given wavepacket function $f$,
take the form
\eqn\psiinit{a^\dagger_{f}= i \int_\sigma dn^\mu
f^\ast\overleftrightarrow{\partial_\mu} \psi\ .}
Specifically, consider the in-state
\eqn\instate{|f_1,\cdots,f_K\rangle=\prod_K a^\dagger_{f_K} |0\rangle\ ,}
and likewise for an out-state with $L$ creation operators.  Our
interest lies in correlators of the form
\eqn\corrstat{\langle f_1,\cdots,f_L| (\calo_{\psi^2 \phi})^N |f_1,\cdots,f_K\rangle\ .}
Let us choose $K$ and $L$ even, with $K+L=2N$, and moreover choose
the in-states such that each pair of ingoing wavepackets $f_{2i-1}$,
$f_{2i}$ overlaps in some definite spacetime region near
$x^\mu=x_i^\mu$, and likewise for pairs of outgoing wavepackets,
but no other pair has substantial overlap in any region of
spacetime.  In that case, \corrstat\ reduces to an expression of the form
%%
%\eqn\psiphicorr{\langle f_1,\cdots,f_L| (\calo_{\psi^2 \phi})^N
%|f_1,\cdots,f_K\rangle \approx C \langle 0| \phi(x_i)\cdots
%\phi(x_N) |0\rangle, }
%%
%where $C$ is a normalization factor depending on the wavepacket
%functions.  Thus, given such states, matrix elements of the
%generalized observables \psf\ reduce to local correlators of the
%field $\phi$.  While the
%operator $\calo_{\psi^2\phi}$ is Poincar\'e invariant, the
%Poincar\'e invariance is effectively spontaneously broken by the
%specification of the states. In such cases, $\calo_{\psi^2 \phi}$
%qualifies as a pseudo-local observable.
%
\eqn\psiphicorr{\langle f_1,\cdots,f_L| (\calo_{\psi^2 \phi})^N
|f_1,\cdots,f_K\rangle \approx C \langle 0| \phi(x_1)\cdots
\phi(x_N) |0\rangle \ . }
 One can thus approximately extract local observables from expectation values of products of $\calo$'s.  In the infinite volume limit there is, however, a subtlety; due to fluctuations, the $\calo$'s are not well defined operators on the Hilbert space of states.  This problem apparently can be suppressed for finite large volume through careful choice of operators.  It does, however, raise possibly fundamental issues, that could be relevant in quantum cosmology, and may have implications for example in the context of interpreting eternal inflation.

More generally, one could also consider the case of a dynamical
metric.  In this situation, one should generalize the states
\instate\ to states solving the Wheeler-DeWitt equation \wdw\
which correspond to incoming (or outgoing) wavepackets coupled to
the metric.  To the extent to which such states can be defined,
one expects to recover a relationship of the form \psiphicorr.

The distinction between the $Z$-model and the $\psi^2\phi$ model
lies in the specific position information being parametrized  in the
operator variables in the $Z$-model, but in the quantum state in the $\psi^2\phi$ model.  In particular, in the $Z$-model we defined a four-parameter family of operators $\calo_\xi$, where for a given choice of state we may dial the parameters $\xi^i$ in order to sample the physics in different regions of the spacetime.  In contrast, we defined only {\it one} operator $\calo_{\psi^2 \phi}$ in the $\psi^2 \phi$ model.  There, in order to sample $\phi$-physics in different spacetime regions, one must adjust the state of the $\psi$-field. Nevertheless, in both models it is the interplay
between the chosen observable and a particular class of quantum
states which leads to localization.

As a final observation, notice that the operator \psf\ can
naturally be added to the lagrangian with a coupling constant to
give an interacting theory.  In this case, we may compute
expectation values of the form \psiphicorr\ by differentiating the
path integral with respect to $\lambda$. More discussion of this
kind of relation between single-integral diffeomorphism-invariant
observables and interaction terms in a lagrangian will be given in
section five, where this will provide part of the connection to
the traditional notion of ``measurement" of observables.

\subsec{String theory and two-dimensional gravity}

The general framework we have described can also be illustrated in
the context of string theory, in which the string is viewed as a
model for two-dimensional gravity.  While there are no propagating
gravitational degrees of freedom in 1+1 dimensions, diffeomorphism
invariance nevertheless plays a crucial role.

To begin, let us recall that, at the perturbative level, string
scattering amplitudes are computed as the expectation values of
 vertex operators
$\calV_i$,
\eqn\stringamp{\langle \prod_i\calV_i \rangle,}
 which are defined as a functional integral over geometries and
fields.  The vertex operators $\calV_i$ should be diffeomorphism
invariant, and in particular typically take the form
\eqn\vertop{\calV_i=\int d^2\sigma {\tilde \calV}_i,}
where ${\tilde \calV}_i$ are densities of the appropriate weight.
Thus, the vertex operators of string theory are
diffeomorphism-invariant observables in the two-dimensional
gravity theory on the worldsheet.

One might ask to what extent the world-sheet fields can be used to
give conditionals defining position,
as in the Z-model of section 4.1.
For example, in the context of
the bosonic string, vertex operators of the form
\eqn\tachvert{{\tilde \calv}= e^{ik\cdot X}}
are commonly considered, where $X^\mu$, $\mu=0,\ldots,D-1$ are the
worldsheet scalar fields.  However, in order for the correlator
\stringamp\ to be well-defined in the critical theory with $D=26$,
the vertex operators \vertop\ must be both diffeomorphism and Weyl
invariant, implying the momenta $k^\mu$ must satisfy the
constraint
$k^2=8$; i.e., they must satisfy
the mass-shell condition of the
target-space tachyon. This means that one cannot treat the
different components of $k^\mu$ as independent integration
variables, and produce sharp gaussians as in \ftgauss.

Relaxation of the condition $k^2=8$ leads to explicit dependence on the conformal part of the metric, $\phi$, where we work in conformal gauge,
\eqn\confgauge{ds^2=e^\phi {\hat g}_{ab} d\sigma^a d\sigma^b\ .}
Here ${\hat g}_{ab}$ is a background metric, which fixes the conformal equivalence class.
Since for the critical string the action is independent of the conformal factor, the expression \stringamp\ is %
no longer well-defined.  This situation changes for the
noncritical theory,
$D\neq26$, where quantum effects induce the Liouville action for
$\phi$,
\eqn\Liou{S_L={25-D\over 48\pi}\int d^2\sigma \sqrt{\tilde g} \left(\hf {\hat g}^{ab} \partial_a\phi \partial^b\phi + {\hat R} \phi\right)\ .}
In general dimension, a matter operator ${\cal W}_i[X]$ of definite conformal
dimension $\Delta_i$ receives a gravitational dressing,
so that, instead of ${\cal W}_i[X]$ itself, it is the operator
\eqn\dressed{{\tilde \calv}_i= e^{\alpha_i \phi} {\cal W}_i\ ,}
which transforms as a density of weight one. Here
\eqn\alphadef{\alpha_i = {25-D\over 12} \left[ 1\pm\sqrt{ 1-D +
24\Delta_i\over 25-D}\right]\ .}
Once again, the dependence of $\alpha_i$ on $k$ restricts our
ability to define gaussians of the $X^\mu$ fields.

In either  critical or non-critical cases, however, it appears
possible in a long-distance approximation to use the operators
$\int {\tilde\calv}_i$ in analogy to the Z-model to specify
location and time information. One could write an expression such
as
\eqn\approxdelta{\int_{-1/L}^{1/L}\prod_\mu dk^\mu e^{ik_\nu\cdot
(X^\nu-\xi^\nu)} e^{\alpha(k)\phi}\ ,}
or,  in the critical case, replace
$\alpha(k)\phi$  by a term proportional to $X^{25}$ and the combined squares of the independent momenta.
For $L\gg 1$, the $k$ dependence in $\alpha(k)$ is small, and on
scales $X\gg L$ one might anticipate this expression approximates
a delta function concentrated at $X^\mu\approx\xi^\mu$, which in turn could
be used to specify worldsheet position.

In the critical case, this can in particular be illustrated by
working about a background corresponding to a string wound on a
non-contractible cycle, of the form $X^0=p \tau, X^1 = w \sigma$.
For $25 > D > 1$, additional subtleties arise as one must deal
with the so-called $c=1$ barrier. Dynamics in this regime is
poorly understood, but it is believed that one encounters a phase
such that the geometry is a branched polymer.  Thus, while the
general framework we describe plays a role here, one won't
necessarily have a phase in which the two-dimensional geometries
have clean semiclassical behavior and permit the existence of
useful clocks. We presume this is a feature unique to
two-dimensional physics, which typically has large fluctuations on
all scales, and based on empirical observation, do not expect such
limitation on our discussion of four-dimensional physics. Indeed,
due to the branched polymer structure (and in contrast to the
higher-dimensional case), it is not even clear what form of local
physics one might wish to recover.

\subsec{Cosmological observables}

Since cosmology is an important domain in which to describe
observation, we briefly comment on how the approach outlined above
may be used to define
relevant observables.  In particular, in the cosmological
context, one is interested in describing observables at different
times in some cosmological evolution.  Objects of particular
interest include correlators of the inflaton and information about
the temperature and geometry of the early universe.

Some of the information of interest requires only locality in
time.  For example, if we are interested in the temperature of the
universe at the end of inflation, we might begin by studying the
energy density at the time when the effective cosmological
constant drops to some level well below the GUT scale.  In the
mini-superspace truncation, we might describe this using as a time
variable the radius of the universe.
This radius is of course not locally defined, but quantities
such as the curvature
are, and allow us to generalize the idea beyond mini-superspace.
 In particular,  in the case where the universe is spatially compact, we may
investigate such quantities through observables of the form
\eqn\cosmorho{\calo_\tau = \int d^4x \sqrt{-g}  O(x)  f_{\tau}(R)  \ ,}
where $O(x)$ is a local scalar operator and $f_{\tau}(R)$ is a
sharply peaked function of the spacetime scalar curvature $R$
with peak near some value $\tau$,
which thus serves as an approximate time label.\foot{More generally, one may wish to use an appropriate {\it spatial average} over the curvature; precise specification of such a prescription is more complicated, but similar to the construction of ``bilocal" operators that will be described below.}
The observable
\cosmorho\ roughly corresponds to the value of the observable $O$
at the given value of $R$.  In states where we expect the universe
to be very homogeneous, there is no need to attempt to resolve
spatial information, or even to localize \cosmorho\ in space.

For example, in the case of the energy density, we might use a
quantity such as
\eqn\locen{O(x)=T^{\alpha \beta} {{\partial_\alpha R
\partial_\beta R} \over {\sqrt {-|\partial R|^2}}}\ ,}
where $T^{\alpha \beta}$ is the stress-energy tensor. Here the
symbol $|\partial R|^2$ denotes the norm of the covector
$\partial_\alpha R$.  In the case mentioned above, the value of
$\tau$ might be chosen to correspond to an effective cosmological
constant at some level below the GUT scale, and the observable
\cosmorho\ then roughly corresponds to the total energy of the
universe at the given value of $R$. Of course, this depends
 on both the total
 volume and on
the energy density.  To recover information about the
energy density (and thus the temperature) alone, we might divide
by, e.g.,
\eqn\cosmoV{\calo_V = \int d^4x \sqrt{-g}  f_{\tau}(R)   \sqrt{-|\partial R|^2}     \ . }
A similar quotient, defined through some limiting procedure, might serve as a useful pseudo-local probe of temperature in cases where the universe is nearly homogeneous, but
is not spatially compact.

In short, one may adapt to our framework the common idea (see,
e.g. \refs{\BryceII}) that one may use the `size' of the universe
to label times when the universe is nearly homogeneous. This idea
has often been implemented in the mini-superspace truncation,
which amounts to using a toy 0+1
model.  In this context, operators analagous to $\calo_\tau$ were studied in greater depth in
 \refs{\QORD\BIX-\MarT}.

We emphasize that by using the local
notion of the spacetime curvature scalar $R$ instead of the
non-local notion of the `size' of the universe, our definition
\cosmorho\ can make sense even in the presence of inhomogeneities
(in which case it merely gives a spatial average of the desired
energy density).  Thus, we expect that $\calo_\tau$ will define an
operator in quantum 3+1 effective gravity.

On the other hand, in the context of homogeneous cosmologies, we
expect information about inflaton correlators to be encoded in
more complicated observables, which are {\it not} of the
single-integral type.  The point is that one needs a means of
specifying the separation between the two operators in a two-point
function in a context where the one-point functions are
independent of position on the homogeneous slice. This suggests
one should build an operator of the form\foot{Bilocal and other diffeomorphism invariant quantities have also been employed in simplicial quantum gravity based on Regge calculus  \refs{\Hamber}, and in two-dimensional gravity in \refs{\Ambjorn}.}

\eqn\cosmoinfl{\calo_{\tau,\Delta} = \int d^4x d^4y\sqrt{-g} \sqrt{-g} f_{\tau,\Delta} (x,y) \phi(x)\phi(y)  \ }
which samples the bi-local operator $\phi(x)\phi(y)$ only when the
two points $x$ and $y$ have some physically specified separation
$\Delta$.  This can be done by, for example, using an operator
$f(x,y)$ whose classical limit is sharply peaked when $x,y$ are
separated by a geodesic of length $\Delta$ lying in the surface in
which the scalar curvature $R(x)$ takes the value $\tau$.  For
example, one might take
\eqn\inflsamp{ f_{\tau,\Delta} (x,y) =   f_{\tau}(R(x)) \   f_{\tau}(R(y)) \   f_{\Delta}(s(x,y))\ ,}
where $f_a(b)$ is the sampling function from \cosmorho\ and
$s(x,y)$ is any functional of the metric which approximates the
geodesic distance between $x$ and $y$ when i) the quantum state is
sufficiently semi-classical and approximates a universe that is
spatially homogeneous in a neighborhood of some spacelike slice
$\Sigma$ and ii) $x$ and $y$ are both located on $\Sigma$.\foot{The
operator $f_{\Delta}(s(x,y))$ may, in turn, be defined at least on
some open set of such auxiliary states by computing the result on
the classical metric corresponding to one such state and then
expanding $f_{\Delta}(s(x,y))$ as a power series in the metric.}
The resulting operators are complicated; we assume a
renormalization scheme for such operators can be specified in a
low-energy effective theory of quantum gravity.

\subsec{General comments}

The examples we have outlined show how relational data may be
encoded in a combination of state and diffeomorphism-invariant
observables, and in particular allow specification of position
information.  Many other examples of these basic principles may be
considered.  In particular, there is no obvious in-principle
obstacle to constructing such operators purely out of
gravitational data, say by constructing objects relating the
values of different curvature invariants.

Note also that a definition of observables, such as that described above,  is useful for
characterizing the physical states
of a theory with dynamical
gravity.  Given a physical state $\Psi$
satisfying the Wheeler-DeWitt equation \wdw, the above observables may be used to formulate projectors onto solutions with
definite attributes.  This follows by virtue of the statement that
an operator of the form
\eqn\project{\delta(\calo - a)\ ,}
(or more precisely a projector onto a spectral interval of $\calo$)
which projects onto states in which $\calo$ takes value $a$,
commutes with the constraints if $\calo$ does.    Thus
combinations of such projectors can be used to specify attributes
of the physical states in terms of values of the  observables.

Such a specification of states is quite similar in spirit to the
conditional probability interpretation, advanced in \Page.
Kucha\v r \Kuchar\  has argued that this suffers from a {\it
reductio ad absurdum}; a counterargument has recently been
proposed in \Dolby.  However, note that the ``projection
operators" of the latter reference do not in fact act as such.  In
contrast, projection operators defined according to \project\ (or
the more precise spectral interval version) are indeed projectors,
and lead to a different approach to defining probabilities.

\newsec{Diffeomorphism-invariant observables and measurement}

\subsec{Measurement: generalities}

The examples of the preceding section have illustrated how certain
``pseudo-local" diffeomorphism-invariant observables reduce to the
usual local observables of quantum field theory.   As we have
seen, this property is critically dependent on the state(s) in
which the observables are evaluated.

Associated
with the usual observables of QFT is a theory of measurement, see
e.g., \refs{\vonN}.  One assumes the existence of an appropriate
 measuring apparatus, whose coupling to the quantum system is
capable of measuring the eigenvalues of the operator in question.
In this section, we describe some aspects of measurement theory for relational observables.

%In the gravitational setting, we have seen that the requirements
%of diffeomorphism invariance can be satisfied by integrating over
%the entire spacetime.
 In the gravitational setting we have seen that, though one must be aware of important infra-red issues,  the requirements
of diffeomorphism invariance can nevertheless be satisfied by integrating over
the entire spacetime.  In order to define localized operators, one
must also include a reference framework.  Specifically, localized
information about some degrees of freedom can be recovered by
constructing operators which explicitly refer both to those
particular degrees of freedom (which we may call the ``target''
degrees of freedom) {\it and} to other dynamical degrees of
freedom; the additional degrees of freedom can specify the
location at which the target degrees of freedom are to be sampled.
In some cases, these additional degrees of freedom might be thought of as providing
an abstract background
of `clocks and rods' against which to localize
the target degrees of freedom,
though of course this background will be dynamic and will be
influenced by the target degrees of freedom.
 Moreover, in any context where one would consider a
local measurement to have taken place (e.g., in a specific
laboratory), it is natural to include degrees of freedom
describing the measuring apparatus, and, in fact, it is natural to
use the apparatus itself to specify the spacetime regime in which
the target degrees of freedom are sampled. Specifically, the
sampling occurs at the location of the apparatus and during the
time interval in which the apparatus is switched on.

This fits with the broader perspective that in a fully quantum
mechanical framework, there should be no sharp distinction between
the observed system and the measuring apparatus -- they are both
quantum systems, with some coupling between them. In this context,
a simple viewpoint is that {\it measurement is correlation with a
subsystem that can be understood as a measuring apparatus:} a
measurement is performed when the system being observed and the
measuring apparatus are allowed to interact, and form correlations
between their degrees of freedom.  This is a general notion for
quantum systems.  One more specifically can speak of a Copenhagen
measurement situation, in which the Copenhagen formulation of
quantum mechanics can be reproduced\foot{For further discussion of
this idea, see {\it e.g.}
\refs{\HaQMC\Omnes\ZII\Joos\ZI-\Schloss}.  In \refs{\HaQMC} such
Copenhagen measurement situations were referred to as ``ideal
measurement situations."}. A measurement framework is Copenhagen
to the extent it can be thought of as describing a quantum system
interacting with a classical measuring device.  Several critical
aspects play a role.  First, the Hilbert space should decompose
into states of the system and states of the measuring device.
Second, the system variables and the corresponding variables of
the measuring device should be exactly correlated, so that the
measurement is good.  Third, the combined
 system should decohere, so that consistent probabilities can be assigned to the different alternative results of
measurements.
Finally, the measuing device should form stable records that are robust against fluctuations and further inspection.  As we will
discuss below, such conditions can be satisfied when the measuring apparatus has a large number of degrees of freedom.

\subsec{Measurement and relational observables}

Although they are non-local, a  connection between measurement and correlation can nevertheless emerge from
a treatment of relational diffeomorphism-invariant observables.
 However, the
correlations we desire will typically arise only in special states
of the system.  This is a standard feature of measurement
situations (see {\it e.g.} \refs{\vonN}), but is especially
prominent here since, as described in section 4, the state plays a
key role in the recovery of the notion of locality itself.

Thus, and in line with the above discussion, a link between
measurement and relational observables arises when specific
conditions hold. The first is that the state and dynamics must
allow an approximate division of the degrees of freedom of the
universe into the measured (target) system and the measuring
device; these may possibly be supplemented by other degrees of
freedom irrelevant to the discussion.  Second, the coupling
between these two systems should be weak, in a sense to be
described shortly. Since we describe the measurement within
effective low-energy gravity, the coupling must be diffeomorphism
invariant. Furthermore, if the effective description of the
coupled system is local, the coupling must provide a term in the
action which is an integral of a local density.  Thus, this
coupling is precisely given by a single-integral observable.

Before proceeding, we pause to clarify one conceptual point.  In
practice, measurement always occurs within some given physical
system.  For example, our laboratories are filled with devices
which, together with their couplings to any target systems, are
described by the standard model of particle physics.  In
particular, the laboratory technician has no freedom to adjust any
coupling constants of the standard model.  However, it is often
useful to give a low-energy effective description of these devices
in which their construction from standard model fields is not
explicit.  Of course, in resonance with our recurring theme, such
an effective description is valid only when the full system (i.e.,
the standard model fields) is in an appropriate state, and
interesting features of the effective description can depend on
the details of the state (e.g., whether the device is ``on'' or
``off'').  This state-dependence gives rise to coupling constants
in an effective description which {\it are} under the control of
the technician.  As a result, measurement theory is typically
discussed in terms of deforming the action of some (typically
uncoupled) system of target and apparatus by introducing some new
coupling between them.  We will pursue this approach below.

Specifically, given an action $S$, let us consider its
perturbation by a single-integral observable $\calo$ of the form
\difop; i.e., we deform the Lagrangian through
\eqn\actpert{ {\cal L} \rightarrow {\cal L}' = {\cal L}+ f\hat \calo\ ,}
where $f$ is a small parameter, and $\calo$ and $\hat \calo$ are
related as in \difop. Such a perturbation of the action leads to a
shift in the inner product \prodstat, inducing new correlations
between the target system and apparatus.

To find this shift, first note that the functional integral in
\prodstat\ will in general be defined over some {\it fixed} range
of parameter time; one then integrates over all geometries
interpolating between the endpoint field configurations in this
parameter time interval.  For example, one may take this parameter
range to be $(0,1)$, and this defines the limits on the integral
determining the action in \prodstat.  In this case, the change of
\prodstat\ under the perturbation \actpert\ is
\eqn\pertprod{\eqalign{\delta \langle {h_2,\phi_2^r}\pip \eta \pip
{h_1,\phi_1^r}\rangle &= if \int_{h_1, \phi^r_1}^{h_2,\phi^r_2}
\cald g \cald \phi^r e^{iS} \int_0^1 dt d^3x\sqrt{-g} \cr
&if\langle {h_2,\phi_2^r} \pip\eta\cohat(t,x) \pip
{h_1,\phi_1^r}\rangle -\delta\left\{ \langle
{h_2,\phi_2^r}\pip\right\}  \pip {h_1,\phi_1^r}\rangle -  \langle
{h_2,\phi_2^r}\pip \delta\left\{\pip
{h_1,\phi_1^r}\rangle\right\}\ ,}} with \eqn\statechangeII{
\delta\left\{ \langle {h_2,\phi_2^r}\pip\right\}    \pip
{h_1,\phi_1^r}\rangle := if \int_{h_1, \phi_1^r}^{h_2,\phi^r_2}
\cald g \cald \phi^r e^{iS} \int_1^\infty dt d^3x\sqrt{-g}
\cohat(t,x) \ ,} \eqn\statechangeI{ \langle {h_2,\phi_2^r}\pip
\delta\left\{  \pip {h_1,\phi_1^r}\rangle \right\}  := if
\int_{h_1, \phi^r_1}^{h_2,\phi_2^r} \cald g \cald \phi^r e^{iS}
\int_{-\infty}^0 dt d^3x\sqrt{-g} \cohat(t,x) \ .} In expression
\statechangeII, the integral is over paths which begin at $t=0$,
advance in $t$ to the far future  and then return to $t=1$.
Expression \statechangeI\ is similar.  The construction is
analogous to that used in the $\langle in | in \rangle$ formalism.

Let us assume that contributions to \statechangeII\ and \statechangeI\ come only from regions far from the Planckian regime.  For example, we expect this to hold for operators $\hat \calo(x,t)$ which, on classical solutions approximate to $|\Psi_1\rangle$, $|\Psi_2\rangle$, happen to be supported in such regions of spacetime.
It is now clear that  \statechangeII\ and \statechangeI\ may be interpreted as changes in the states $\eta \pip h_1, \phi_1^r\rangle$ and   $\eta \pip h_2, \phi_2^r\rangle$ when these states are held fixed at, respectively, late and early times, perhaps as they emerge from a region of Planck scale physics\foot{We make the implicit assumption that, for states of interest, regimes of $(t,x)$ contributing to \statechangeII\ and \statechangeI\ are not separated by intervening Planck-scale physics. This
in particular requires exclusion of evaporating black holes and phenomena such as bouncing universes.}.

More generally, we can superpose the quantities \pertprod\ to find the change in the inner product between two arbitrary auxiliary states, $\langle \Psi_2 \pip \eta \pip  \Psi_1\rangle$.
With the above understanding of boundary conditions, we may describe this as the change in the physical inner product $\langle \Psi_2 | \Psi_1 \rangle$.  That is, we define $\delta \langle \Psi_2 | \Psi_1 \rangle$ to be
$\delta \langle \Psi_2 \pip \eta \pip \Psi_1 \rangle$ where the auxiliary states $\pip \Psi_1 \rangle,\pip \Psi_2 \rangle$ are chosen so that  $\delta\left\{ \langle {\Psi_2}\pip\right\}    \pip  {\Psi_1}\rangle $ and  $ \langle {\Psi_2}\pip  \delta\left\{  \pip  {\Psi_1}\rangle \right\}$ are both small; we make no definition of $\delta \langle \Psi_2|\Psi_1 \rangle$ when such a choice is not possible.

Thus, we have derived a diffeomorphism-invariant version of the Schwinger variational principle \refs{\Schwing,\BryceBook} relating the change in this inner product to the matrix element of our diffeomorphism-invariant observable,
\eqn\statech{\delta\langle \Psi_2 | \Psi_1\rangle = \langle \Psi_2|\calo | \Psi_1\rangle\ .}

 Said differently, we take
 the initial and final states $|\Psi_i\rangle$, $i=1,2$,  to be specified in terms of data associated with a region undisturbed by the interaction $\cohat(x,t)$, where possible.  This data is
encoded through the choice of auxiliary states $\pip \Psi_i
\rangle$, for which $|\Psi_i \rangle = \eta \pip \Psi_i \rangle$
and for which $\delta  | \Psi_i \rangle$ as defined above is
small. A more complete way of stating this is to say that we start
with a notion of asymptotic physical states, in some basis, in
both past and future, analogously to what we do in the LSZ
framework in field theory.  We assume that the perturbation
\actpert\ has negligible effect on the form of the ``in" states in
the past, or on the form of the ``out" states in the future.  Of
course, complete specification of the states involves physics at
the Planck scale, so here we must make the assumption (which we
consider reasonable, based on simple examples) that we are working
in a sufficiently semiclassical regime that we can specify the
states in the effective theory and that the operator in question
in effect turns off in the past and future.

Our basic picture is then that the left side of eq.~\statech\ can,
in these circumstances, be related to the result of a measuring
process; this then
provides a measurement interpretation of the matrix element on
the right side of this equation.   Specifically, start with the
assumption that the state is such that there is a clean division
between
target system and measuring apparatus, with only a weak
interaction between them. For example, we might consider the
situation where the target system corresponds to one of the
fields, which we call $\phi(x)$.  A concrete example to bear in
mind is that the field describing the system might be, {\it e.g.}
the muon field, whereas the measuring apparatus is constructed
from electrons, protons, {\it etc.} The Wheeler-DeWitt
wavefunction should be linear combinations of auxiliary states of the form
 \eqn\prodstates{ \pip \Psi_A \rangle = \pip \alpha, a, h_A \rangle = |\alpha\rangle_\phi |a\rangle_m |
h_A\rangle_h
\ ,}
where the factors are states $| \alpha\rangle_\phi$ of the
target system, $| a\rangle_m$ of the measuring device,
and $|
h_A\rangle_h$ of the metric (and possibly other degrees of
freedom).  The state of the metric then becomes correlated to that of the
system and measuring device through the Wheeler-DeWitt equation
\wdw; i.e., in the corresponding physical state $\eta \pip \Psi_A \rangle$.  The interaction between the system and measuring device
will typically be of the form of a single-integral
diffeomorphism-invariant operator,
\eqn\sysmeasint{S_i=f\calo=f\int d^4 x \sqrt{-g} O(\phi(x)) m(x)}
where $O(\phi(x))$ is a local operator constructed from the field
$\phi$ and $m(x)$ is an operator acting on the state of the
measuring device.

Working about a background which is sufficiently semi-classical (which presumably requires gravity to be weakly coupled),
 the inner product \prodstat\ of states of the form  \prodstates\ is approximated by matrix elements of a so-called ``deparameterized theory,'' in which the constraints have been solved and one finds an ``external time'' which plays the same role as time in ordinary quantum field theory (or, for that matter, in non-relativistic quantum mechanics).  This external time may arise either from clock degrees of freedom in the measuring apparatus, or from the metric background. Work along these lines has a long history; see, e.g., \refs{\BryceII,\BanksTCP,\Rubakov,\Kiefer}, and in particular \refs{\MarT} for a careful discussion in terms of pseudo-local observables (in the 0+1 context).

 If $U$ is the evolution operator of the target system and measuring device in the deparameterized theory, the relation takes the form
\eqn\approxprod{\langle \Psi_{B}| \Psi_{A}\rangle \approx \langle \beta, b|U| \alpha, a \rangle e^{iS[g_{cl}] }\ , }
where the states on the right hand side lie in the deparameterized theory (so that the clock degrees of freedom no longer appear in the state).
 The assumption that the system and measuring device are weakly coupled justifies the approximation in \pertprod\ of truncating to linear order in the coupling $f$, so that \statech\ may be written
\eqn\intpict{\langle\beta, b | (U-1) |\alpha, a \rangle = i f \int d^4 x \sqrt{-g_{cl}} \langle \beta| O(\phi(x))|\alpha\rangle \langle b|m(x)|a\rangle + \calo(f^2)\ ,}
which agrees with the interaction typically used to discuss measurement of the local field theory observable $O(\phi)$ in the spacetime region in which the device $m(x)$ is active.  Physically, the measurement proceeds through the establishment of correlations of the $\phi$ system with the measuring device.  If the device is sufficiently classical, a Copenhagen measurement results.

While we have outlined the connection to measurement as if the
degrees of freedom of the measured system are a different type of
field than those of the
target system, the discussion generalizes readily to the
situation where both measured system and measuring device have the
same constituents, {\it e.g.} electrons. In this case the
decomposition \prodstates\ corresponds to
factoring the auxiliary Hilbert space into a product of  Hilbert
spaces corresponding to distinct degrees of freedom of the
electron field, and likewise the two operators in \sysmeasint\ are
operators that act on the two different sets of degrees of
freedom.  (The general interaction/single-integral observable will
be a sum of such terms.)

In short, the fact that \intpict\ approximates \statech\ makes it
clear that, just as in more familiar (e.g., \refs{\vonN})
discussions of measurement, when the states and observables are of
a specific form, measuring devices become correlated with states
of the target system in such a way that the outcome of the
measurement is given by the matrix elements of the pseudo-local
observable $\calo$.  As in the case of measurement theory in the
presence of an external time, one may also ask about the degree to
which such correlations may be viewed as {\it Copenhagen}
measurements; i.e., measurements to which the Copenhagen
interpretation of quantum mechanics can be consistently applied.
This question is examined in the following subsection, and again
in section 6, where constraints imposed by gravity are discussed.

\subsec{The Copenhagen measurement approximation; large $N$}

Having described measurements of (single-integral)
diffeomorphism-invariant observables, one may also ask to what
extent such measurements can approximate Copenhagen measurements.
In particular, we expect to precisely recover the needed
properties of {\it decoherence} and {\it stability} only in the case
of measuring devices comprised of infinitely many degrees of
freedom (here we may also wish to include other variables
describing the environment as part of the measuring device; these
can be important for ensuring decoherence).  In a later section,
we will discuss gravitational constraints on numbers of degrees of
freedom, but for the moment let us
consider more generally the limitations imposed if the number of
degrees of freedom of a measuring device is finite. Thus,
diffeomorphism-invariance will not play a direct role in the
discussion below.

For illustration, we consider the Coleman-Hepp model
\refs{\Hepp,\Bell}; for other examples making use of an
``environment,'' see e.g. \refs{\Omnes\Joos\ZI\ZII-\Schloss}. This
is a quantum-mechanical model for a device that measures the state
of a two-state quantum system, for example the spin of an
electron.  The measuring device consists of $N$ two-state spins.
Let the states of the ``electron" be denoted
$|+\rangle,|-\rangle$, and
 states of the measuring device be of the form
$|\uparrow\downarrow\cdots\uparrow\rangle$.  An explicit
hamiltonian can be written down, but all we need is the result of
its evolution:  a general state of the two-state system (combined
with some specific initial state for the measuring device) evolves
into a perfectly correlated state,
\eqn\cohepevol{\alpha|+\rangle + \beta |-\rangle \rightarrow \alpha|+\rangle|\uparrow^N\rangle + \beta |-\rangle|\downarrow^N\rangle\ .}
Thus the system variables and measurement variables are indeed perfectly correlated.

The limitations arising from finite number of degrees of freedom
are manifest in the conditions of decoherence and stability.
For the state on the left hand side of \cohepevol, interference
effects are important for computing the expectation values of many
operators, such as, e.g. $\sigma_x$ or $\sigma_y$.  In the
docoherent histories formulations of quantum mechanics (see, e.g.
\HaQMC\ and references therein), the corresponding statement is that a
typical set of alternative histories will not decohere.  Of
course, the state on the right-hand side of \cohepevol\ is also a
quantum state for which interference can be measured, but as $N$
grows this becomes increasingly difficult, as the phase
information becomes distributed over a larger number of degrees of
freedom. Thus as $N$ gets large,
interference effects are supressed for operators involving only a
finite number of spins, or, equivalently, typical sets of alternative histories decohere. To make this more precise, the only
operators that are sensitive to interference between the two
components of the composite state \cohepevol\ are composite
operators that act on {\it all} of the $N+1$ degrees of freedom:
\eqn\coh{\langle \uparrow^N | \langle+|O |-\rangle|\downarrow^N\rangle \neq 0}
only for an operator $O$ that flips all the spins, {\it e.g.}
\eqn\egop{O= \sigma_y^{\rm system} \prod_{i=1}^N \sigma_y^i\ .}
In the ``classical" limit of $N\rightarrow\infty$, no operator $O$
acting on a finite number of spins is sensitive to this
interference. Likewise, stability improves with increasing $N$.
Real systems are difficult to isolate, and generic small
perturbation terms in the hamiltonian, {\it e.g.} due to
interactions with the environment or other effects, will typically
randomly flip individual spins. However, if the probability to
flip a single spin in a given time interval is $\gamma < 1$, the
probability to flip more than half the spins of the measuring
device, and thus spoil the measurement, is $\gamma^{N/2}$ which
vanishes as $N\rightarrow\infty$.

We see that at infinite $N$ the expected classical behavior is
recovered, but for finite $N$
there are limitations on the extent to which one
can achieve a classical measurement. Put more descriptively, if
we make an
 observation of one alternative, but then via a quantum or other
fluctuation,
 our brain transitions into a state corresponding to a different
alternative, we ultimately reach a different conclusion about the
outcome of the measurement. Such fluctuations
 are always in principle possible for finite systems. This
suggests that any such measurement has an intrinisic uncertainty
that falls exponentially with the number of degrees of freedom of
the measuring apparatus,
\eqn\uncert{\Delta \approx e^{-cN}\ ,}
where the constant $c$ depends on the details of the apparatus.
Similarly, quantum interference effects mean that the measurement
will fail to be Copenhagen also at order
$e^{-cN}$.  Ref.~\BanksWR\ has previously emphasized the importance of uncertainties of this magnitude, and made a similar estimate from quantum tunneling.
We will discuss gravitational restrictions on this number of
degrees of freedom in the next section.

Finally, a remaining source of uncertainty is the limited
resolution provided by a system with a finite number of bits.
Whenever one attempts to measure what might be a continuous
parameter, using an $N$-bit device, one expects that the result
stored has an uncertainty of the form $\Delta \sim
2^{-N}$.

To summarize this section, we see that in cases where there is
a decomposition into target system and measuring device degrees
of freedom, along with remaining metric and other degrees of
freedom, such that interactions between the system, measuring
device, and other degrees of freedom are weak, and such that the the
measuring device is well approximated as a classical measuring
device,
one can recover measurements of a quantum system, with
the results corresponding to matrix elements of
appropriate pseudo-local
diffeomorphism-invariant observables. In such circumstances relational observables can be given a clear interpretation in
terms of measurement, but such an interpretation does not follow
in the case of more general dynamics and states.

\newsec{Observables: limitations}

The preceding sections have described how useful
diffeomorphism-invariant observables may be constructed in an
effective low-energy quantum theory of gravity, and
argued that, in some circumstances, these observables reduce to
local observables of standard quantum field theory (QFT).
However, we also
found limitations on recovering QFT observables from our
diffeomorphism-invariant observables. Some of these arise from
basic quantum properties, and were touched upon in section 4.1.
However, it appears that additional limitations arise when we take
into account the coupling to a dynamical metric.  In this section
we examine both kinds of constraints more completely, and discuss
their possible role as fundamental limitations on the structure of
physical theories.

\subsec{Example of the $Z$ model}

We begin by investigating constraints on observables in the context of the $Z$ model.  Recall that we argued that the diffeomorphism-invariant observables of the model approximately reproduce the local observables of QFT, but with limitations on the spatial resolution of the QFT operators.  These limitations
stem from two sources.  First, the position resolution of the operator in \opdef\ is limited by the value of $\sigma$; recall that a non-zero $\sigma$ is required to regularize the operator.    Second, when we use the variables $\xi$ to fix the spatial coordinates, we find that fluctuations become strong and we lose control when the resulting separation between two operators is too small.
The resolution $\Delta x$ is limited by the large fluctuations \correstone\ of the $Z$ fields at small separations $|x_1-x_2|$.
Together, these two features limit the
resolution
at which we can independently measure separate degrees of
freedom of the field $\phi$.  Specifically, the physics of two separate local operators at $x_1, x_2$ is reproduced only when the separation between the operators satisfies
\eqn\xlimits{|x_1-x_2| \roughly> {\rm Max} \left({\sigma\over \lambda}, {1 \over \sigma} , {1\over \lambda |x_1-x_2|}\right). }
Here the first condition follows from
\xgauss\ and the fact that we wish to separately resolve the two observables, while the second condition follows from \correstone\ and the third from
\corresttwo\.  Note that the first two conditions imply the third, so that \corresttwo\ does not play a key role in the discussion.
In order for
fluctuations to be under control, we find from \correstone\ that
the dominant contribution to this uncertainty
must be that of $\sigma/\lambda$.

In order to minimize this uncertainty,
one wishes to maximize the $Z$-field gradient $\lambda$.  In
doing so, however, we should bear in mind that we are ultimately
working in a field theory with a cutoff.  The maximum value for
the field momentum is thus determined by the cutoff as
$\lambda\roughly<\Lambda^2$. Moreover, the minimum value of
$|x_1-x_2|$ should likewise be $1/\Lambda$, and \correstone\ thus
imposes the constraint $\sigma\roughly> \Lambda$.  The net
result is that the fundamental limitation on the resolution is
given in terms of the cutoff by
\eqn\deltaxmin{\Delta x\roughly> {1\over \Lambda}\ ,}
as discussed in section 4.1, and as expected.

Thus, under purely field-theoretic considerations, we might expect
to be able to choose a resolution
limited only by that of the cutoff of the field theory
used to specify location.  Without gravity,
there need not be a fundamental limitation on the size of this
cutoff. Including gravity, one might expect that the Planck scale
serves as a limitation on resolution.  However, the inclusion of
gravity also leads to additional constraints, to which we now
turn.

Suppose that we couple the Z-model to the gravitational field.
 The $Z$ fields serve as a source of
 gravity through its stress tensor,
\eqn\Zstress{T_{\mu\nu}= \hf \left[\nabla_\mu Z^i \nabla_\nu Z^i - \hf g_{\mu\nu} (\nabla Z^i)^2\right]\ .}
Consider attempting to define  observables throughout a spacetime region $\Omega$, choosing a state such that \Zvev\ holds throughout the region.  This means that the stress tensor has size
\eqn\Tsize{\langle T_{\mu\nu}\rangle \propto \lambda^2}
throughout $\Omega$.  If $R$ is the linear size of the region,
then the entire system undergoes gravitational collapse and our
framework for defining observables breaks down if
\eqn\Rmax{\lambda^2 R^3 \roughly> R M_p^2\ .}
This simplifies to the bound
\eqn\maxsize{R\lambda\roughly<M_p\ }
relating $R$ and $\lambda$.

For example, suppose that we wish to provide $Z$ fields
which ``instrument" the region $\Omega$ at the maximum
resolution $1/\Lambda$ allowed in the cutoff theory.  In this
case, we find the bound
\eqn\Rbound{R\roughly< {M_P\over \Lambda^2}}
for the maximum sized region, given the resolution $\Lambda$. A bound of this form on the domain of validity of effective field theory has previously been proposed  by Cohen, Kaplan, and Nelson
in \refs{\CKN}.

There is a similar bound involving
pairs of operators.  In particular, consider a correlation
function of $\calo_\xi$'s of the form \corrdef.  Suppose that we
want each of the positions to be resolved at a maximum resolution
$1/\Lambda$. In particular, this means that each of the operators
has an energy of order $\Lambda$.  Thus for two operators with a
separation $|x_1-x_2|$,  gravity will become strong and our
description of the observables will break down for
\eqn\localbd{\Lambda \roughly > |x_1-x_2| M_P^2\ .}
In fact, this bound is implied by
what was termed the ``locality bound'' in
\refs{\GiLione,\GiLitwo}.

Within the context of a given effective field theory, the bound
\localbd\ is trivially satisfied for $\Lambda < M_p$, as it can be
violated only for $|x_1 - x_2 | < 1/\Lambda$.  However, boost
invariance of the underlying theory indicates that we can create a
particle with ultra-planckian momentum by performing a
sufficiently large boost on a state with sub-planckian momentum,
and one might correspondingly expect one could describe
single-particle states with resolutions $1/\Lambda< 1/M_p$ using
such a boost.  Suppose we view such a state as being created by a
pseudo-local operator.  One can then ask if there is any
in-principle obstacle to such a construction.  The locality
bound\refs{\GiLione,\GiLitwo} states that there should be, since,
if two such operators exceed the bound \localbd, strong
quantum-gravitational backreaction cannot be ignored.

\subsec{General discussion}

While the
above bounds were illustrated using our model for
observables and measurements arising from our $Z$ fields,
one expects them to reflect a quite general situation.
To see this, note first
 that constructing any kind of field configuration -- whether
from the metric, matter, or other fields -- that has a ``resolving
power" $1/\Lambda$, requires working with fields with momenta
$\sim\Lambda$, and hence corresponding energies.  If we want to
construct a ``grid" from these fields, capable of this resolution
throughout a region of size $R$, the energy of the ``grid" is of
order $\Lambda (\Lambda R)^3$. The constraint that the size of the
region be greater than the Schwarzschild radius is thus the bound
of \CKN,
\eqn\genbd{M_p^2 R\roughly > \Lambda (\Lambda R)^3\ ,}
or \Rbound.

Note that this bound is surprisingly strong.  If, for example, we want to ``instrument" a region with fields capable of resolving degrees of freedom at the scale $TeV^{-1}$ throughout the region, the maximum size region has size
\eqn\tevreg{R\sim {M_P\over TeV^2}\ ,}
or in other words, $R\sim 1 mm$ !  This is not a constraint on a given single (or several) particle state in a region, which can be measured with a much smaller resolution; in practice we do so with larger detectors.  But we cannot measure all of the degrees of freedom at $TeV^{-1}$ resolution in a region larger than given by the bound, at least without accounting for black hole formation and the degrees of freedom of gravity at the Planck scale.

Likewise, merely making two measurements in a given region, each
with resolution $1/\Lambda$, involves energies $\Lambda$. Absence
of gravitational collapse thus means that the separation of the
measurements must be greater than the corresponding Schwarzschild
radius, giving the locality bound constraint \localbd.

\subsec{Fundamental limitations on physics?}

We finish this discussion on limitations to measurement by
exploring its consequences for fundamental physics.  One might
take the viewpoint that the constraints of this section simply
arise for the kind of observables that we have described and are
not fundamental constraints on the underlying physics.  However,
it is quite plausible that the approach we have outlined is general enough to yield the most general
observables in a theory with dynamical gravity; it is
not apparent that one can find {\it other} independent constructions of
diffeomorphism invariant operators that can play the role of
observables, much less ones that reduce to QFT observables in the
appropriate approximations. So a natural conjecture is that {\it
all}
observables relevant to the description of local physics
 in a theory with dynamical gravity
 arise from the kinds of observables
that
we have described.

Whether or not this is true, it suggests an even more interesting conclusion.  For example, consider the bound \Rbound\ that says
there is no way to simultaneously measure all of the field theory
degrees of freedom at a resolution $1/\Lambda$ in a region of size
larger than given by \Rbound, using only degrees of freedom
inside the region. One might say that these degrees of freedom
``exist," but simply can't all be described by observables and/or
measured. But an alternative arises if we take
a viewpoint which follows from the principle of parsimony: that
which can't be measured has no existence in physics; physics
should be limited to describing only degrees of freedom that are
at least in principle observable.
Such a viewpoint was useful in the original formulation of
quantum mechanics.  If this principle holds here, one reaches the
conclusion that the maximum number of degrees of freedom
within a cube of size $R^3$ is
\eqn\maxdof{N(R)\sim (M_pR)^{3/2}\ .}
More precisely, this is a proposal for a bound on the number of
states with a non-gravitational quantum field theoretic
description; such a bound was explored in \CKN\ and earlier
noted by 't Hooft\tHooftGX. It is certainly possible that with inclusion of
gravitational degrees of freedom and proper treatment of their
dynamics, and of corresponding observables, a region of size $R$
can support more degrees of freedom.
For example we would not be surprised to find the upper bound
\eqn\bek{N_{BH}(R)\sim (M_PR)^2}
corresponding to the
Bekenstein-Hawking entropy of a black hole, arising from such an
analysis.  Indeed, \refs{\StiffStars} has even argued that \bek\ can be reached through an appropriate choice of equation of state.

Likewise, from the bound \localbd, one
would conclude that there is no sense in which two independent
degrees of freedom with resolution $1/\Lambda$ exist at relative
separations less than given by \localbd.
It was argued in \GiLitwo\  that such logic leads to a loophole in
 Hawking's original argument \Hawkunc\ for information
 destruction by black holes.

When combined with the discussion of limitations from finite
measuring apparatuses of section five, such arguments for
limitations on number of degrees of freedom in a finite region (or
closed universe) indicate an intrinsic uncertainty in measurement.
Such arguments have particular force in de Sitter space, as
described in \BanksWR, which is commonly believed to have only
finitely many degrees of freedom \refs{\Bankslittle,\Fischler}
corresponding to its finite entropy. In particular, if we work
within a region of size $R$ which has a bounded number of degrees
of freedom $N(R)$, then amplitudes that can be measured by devices
constructed in this region have an intrinsic uncertainty of the
form \uncert. This represents an intrinsic uncertainty or
imprecision above and beyond the usual uncertainties arising from
quantum dynamics alone.  One might draw from this this the
conclusion \BanksWR\ that a single mathematically precise theory
of de Sitter space does not exist.  We consider as an alternative
an analogy to quantum mechanics:  once the inevitable uncertainty
in momentum and position was discovered, the relevant question is
what quantity can be precisely predicted, and the answer  is the
wavefunction.  This begs the question:  what is the analogous
fundamental mathematical construction in the present context?

The reasoning we have outlined suggests the outline of a ``first
principles" approach, in analogy with the well-known ``Heisenberg
microscope discussion,'' to understanding the radical thinning of
degrees of freedom that is believed to occur in quantum gravity --
a crucial aspect of the putative holographic principle.  In short,
by the above logic, what can't be observed doesn't exist, and
gravitational dynamics puts unexpectedly strong constraints on
what can be observed.  If this is the case, a very important
question is to come up with a description of the degrees of
freedom and dynamics that do exist, respecting these various
non-local constraints.  We expect this description to look nothing
like local field theory in spacetime; ordinary local quantum field
theory only emerges as an approximation to this underlying
dynamics.

\newsec{Discussion and conclusion}

This paper has addressed the construction and interpretation of diffeomorphism-invariant observables of effective quantum gravity.  In particular, we study operators constructed via integrals, in analogy to the construction of gauge-invariant observables in Yang-Mills theory via traces.  A particularly important class of such operators are the ``pseudo-local" operators, which in certain circumstances reduce to the local observables of field theory.  This happens only in certain states, and the information about location is encoded in the interplay of the operator relative to the state.   Moreover, locality is only recovered in an approximation, and is in general spoiled by both quantum and gravitational effects.  Thus locality is both relative and approximate.

Though single-integral pseudo-local observables experience fluctuations that grow with the infra-red cut-off, for appropriate such operators (e.g., ${\calo}_\xi$ in the $Z$-model) this volume divergence appears with an exponentially small pre-factor.  Thus, in a universe of moderate volume, the effect of such fluctuations can remain small.  Nevertheless, it would be very interesting to understand whether  proper relational observables
can be defined  in the infinite volume limit.  Of course, in this limit other observables exist: the S-matrix.  The relationship between relational observables and the S-matrix is an interesting question for further exploration.
This issue may also have interesting implications for universes with a long period of rapid growth, and in particular for eternal inflation scenarios.

The outline of a theory of measurement for these operators has
also been presented.  This theory respects the idea that there
should be no fundamental separation between the measuring device
and the system being measured. This theory is inherently
incomplete:  we can only explain how to relate matrix elements of
diffeomorphism-invariant observables to results of measurement for
{\it certain observables} and in {\it certain states}.  In
particular, a necessary condition for our discussion of
measurement is the emergence of an appropriate semi-classical
limit.

The further limitations that arise in the treatment of these observables may also represent intrinsic limitations on local physics.
 In particular, these include the statements that spatial resolution in a given region is limited by a lower bound that grows with the size of the region, and that two (or more) particles can only be measured at increasingly fine resolution if their separation increases.  Moreover, these statements also suggest that the number of local quantum degrees of freedom in a finite-sized region is finite.  Combined with the present discussion of measurement, this suggests an intrinsic uncertainty in measurements, above and beyond that of quantum mechanics.

A complete identification of the observables of quantum gravity clearly requires the  full framework of underlying quantum gravitational theory.  We expect that there will continue to be relational observables in this context.  If this is a theory of extended objects, such as strings and branes, this may suggest additional limitations on locality.

Note that our expressions for diffeomorphism-invariant and relational observables bear some formal similarity to observables constructed in non-commutative theories \refs{\GHI} and in open string field theory \refs{\HaIt}.  In particular, the latter take the form
\eqn\SFTobs{\int V\left({\pi \over 2}\right) A\ ,}
where $A$ is the open string field and $V$ is an on-shell closed string vertex operator.  These share the feature that they involve an integral of a product of fields that gives an invariant.  It may be that ultimately similar observables will be discovered in closed string theories, and reduce, in the effective gravity limit, to the kinds of observables we have described in this paper.

The present paper at best only outlines some of the boundaries of
our knowledge of non-perturbative quantum gravity.  However, even
this seems a useful enterprise, and the above limitations support
the statement that these boundaries reach to distances far larger
than the Planck length.

\bigskip\bigskip\centerline{{\bf Acknowledgments}}\nobreak

We have greatly benefited from several conversations with T.
Banks, and acknowledge useful discussions with D. Gross, K.
Kucha\v{r}, and G. Moore.  We thank Jorma Louko and Domenico
Giulini for help in locating references, and S. Hossenfelder for
comments on a draft of the paper. S.B.G. was supported in part by
Department of Energy under Contract DE-FG02-91ER40618, while D.M.
was supported in part by NSF grant PHY03-54978 and J.B.H. was
supported in part by NSF grant PHY02-44764. The authors were also
supported in part by funds from the University of California.

%\appendix{A}{Appendix title}

\listrefs
\end